\documentclass[twocolumn,amsmath,amssymb,superscriptaddress,aps,pra,floatfix]{revtex4-1}

\newcommand{\figRef}{Fig.~}
\newcommand{\eqRef}{Eq.~}

\newcommand{\settitle}{\@maketitle}
\usepackage{mathtools}
\usepackage{braket}
\usepackage{gensymb}
\usepackage{hyperref}
\usepackage{xcolor}
\usepackage{siunitx}
\usepackage{nicefrac}

\begin{document}

\title{Coherent characterisation of a single molecule in a photonic black box}

\author{Sebastien Boissier}
\affiliation{Centre for Cold Matter, Blackett Laboratory, Imperial College London, Prince Consort Road, SW7 2AZ, London, United Kingdom}
\author{Ross C. Schofield}
\affiliation{Centre for Cold Matter, Blackett Laboratory, Imperial College London, Prince Consort Road, SW7 2AZ, London, United Kingdom}
\author{Lin Jin}
\affiliation{Physikalisches Institut, Westf\"{a}lische Wilhelms, Universit\"{a}t M\"{u}nster, Heisenbergstrasse 11, 48149 M\"{u}nster, Germany}
\author{Anna Ovvyan}
\affiliation{Physikalisches Institut, Westf\"{a}lische Wilhelms, Universit\"{a}t M\"{u}nster, Heisenbergstrasse 11, 48149 M\"{u}nster, Germany}
\author{Salahuddin Nur}
\affiliation{Centre for Cold Matter, Blackett Laboratory, Imperial College London, Prince Consort Road, SW7 2AZ, London, United Kingdom}
\author{Frank H. L. Koppens}
\affiliation{ICFO -- Institut de Ciencies Fotoniques, The Barcelona Institute of Science and Technology, 08860 Castelldefels (Barcelona), Spain}
\author{Costanza Toninelli}
\affiliation{LENS and CNR-INO, Via Nello Carrara 1, 50019 Sesto Fiorentino (FI), Italy}
\author{Wolfram H. P. Pernice}
\affiliation{Physikalisches Institut, Westf\"{a}lische Wilhelms, Universit\"{a}t M\"{u}nster, Heisenbergstrasse 11, 48149 M\"{u}nster, Germany}
\author{Kyle D. Major}
\affiliation{Centre for Cold Matter, Blackett Laboratory, Imperial College London, Prince Consort Road, SW7 2AZ, London, United Kingdom}
\author{E. A. Hinds}
\affiliation{Centre for Cold Matter, Blackett Laboratory, Imperial College London, Prince Consort Road, SW7 2AZ, London, United Kingdom}
\author{Alex S. Clark}
\email[Email: ]{alex.clark@imperial.ac.uk}
\affiliation{Centre for Cold Matter, Blackett Laboratory, Imperial College London, Prince Consort Road, SW7 2AZ, London, United Kingdom}
	
\date{\today}
	
\begin{abstract}
Extinction spectroscopy is a powerful tool for demonstrating the coupling of a single quantum emitter to a photonic structure. However, it can be challenging in all but the simplest of geometries to deduce an accurate value of the coupling efficiency from the measured spectrum. Here we develop a theoretical framework to deduce the coupling efficiency from the measured transmission and reflection spectra without precise knowledge of the photonic environment. We then consider the case of a waveguide interrupted by a transverse cut in which an emitter is placed. We apply that theory to a silicon nitride waveguide interrupted by a gap filled with anthracene that is doped with dibenzoterrylene molecules. We describe the fabrication of these devices, and experimentally characterise the waveguide coupling of a single molecule in the gap.
\end{abstract}

\maketitle
	
\section*{Introduction} \label{intro}

Integrated photonic devices have allowed rapid progress to be made in applications such as quantum sensing \cite{Crespi2012,Ono2019}, quantum simulation \cite{Sparrow2018}, and quantum information processing \cite{Qiang2018}. However, the photon sources used in such devices are usually based on probabilistic nonlinear processes. A deterministic photon source would be more useful and single quantum emitters such as quantum dots \cite{Lodahl2015}, defect centres in crystalline materials \cite{Aharonovich2016}, and single organic molecules \cite{Lounis2000} have shown great promise in this regard. A single emitter coupled to an integrated photonic structure can act as a deterministic photon source and can also be used to build photon--photon interactions at the heart of a number of optical quantum computing schemes \cite{Hwang2011a,Javadi2015}.

Typically, the coupling of a quantum emitter to a single-mode fiber or waveguide is quantified by carefully accounting for losses though all elements of the optical setup \cite{Wang2019}. Another method is to compare the lifetimes of two similar emitters, one of which is not coupled to the photonic structure, and to use the Purcell effect to determine the coupling \cite{Arcari2014}. A third approach, known as extinction spectroscopy, relies on the interference between a continuous-wave laser and the resonance fluorescence of the emitter \cite{Javadi2015,Turschmann2019}. This interference affects the amplitude \cite{Gerhardt2007a}, phase \cite{Pototschnig2011} and photon statistics \cite{Foster2019} of the transmitted and reflected fields. Exploration of this phenomenon has led to the demonstration of single emitters as optical transistors \cite{Hwang2009}, phase switches \cite{Tiecke2014} and quantum memories \cite{Bhaskar2020}.
Extinction spectroscopy has been described in a number of settings, including in free space \cite{Zumofen2008, Wrigge2008}, with continuous waveguides \cite{Turschmann2019} and with cavities \cite{Javadi2015, Sipahigil2016, Wang2018a}. Here, we expand the theory to describe an emitter placed in a photonic environment for which we cannot use modal decomposition to find a limited number of relevant modes. We only require that the environment be passive and linear and that the coupling of the emitter to the photonic reservoir be Markovian. We consider an arrangement where two guiding structures are used as input-output ports to the photonic structure, and derive general results for the reflection and transmission spectra as a function of coupling efficiency.

We then apply this result to the characterisation of a single dibenzoterrylene (DBT) molecule coupled to a silicon nitride waveguide. Polycyclic aromatic hydrocarbons were among the first solid-state quantum emitters to be studied \cite{Moerner1989, Brunel1999a}, and have now become a significant alternative to other emitters \cite{Rattenbacher2019, Wang2018a}. Organic emitters have typically been coupled to inorganic photonic structures through  evanescent coupling \cite{Turschmann2017,Lombardi2018,Grandi2019,Rattenbacher2019}. However, the coupling is strongest at the maximum of the field and this motivates the geometry we consider here, where we investigate a waveguide structure interrupted by a microfluidic channel. We demonstrate that the channel can be filled at an elevated temperature by molten anthracene doped with DBT, and that the DBT can have narrow resonances in the vicinity of the waveguide when cooled to cryogenic temperatures. We use extinction spectroscopy to characterise the coupling of the emitters to the waveguide, and because our photonic structure does not admit well-defined optical modes, we use our general theory to fit the transmission spectrum and quantify the coupling. Finally, we compare that measured coupling with the coupling expected from numerical simulations.

\section*{Results}


\begin{figure}
	\begin{center}
		\includegraphics[width=1\columnwidth]{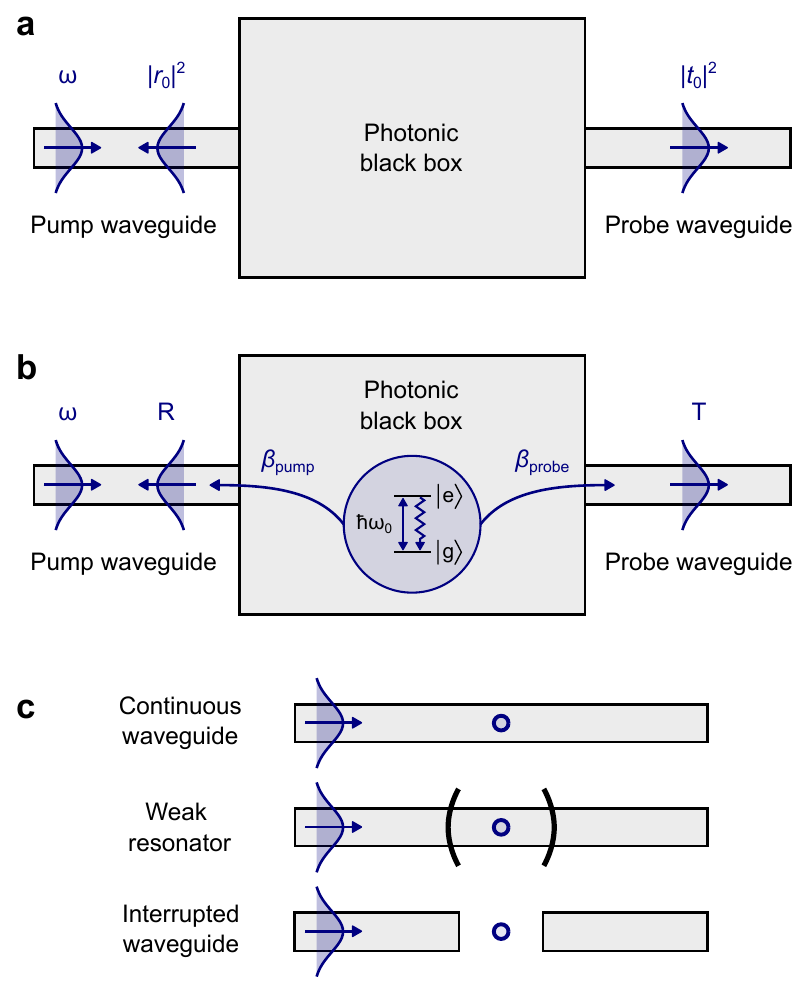}
	\end{center}
	\caption{\textbf{Schematic representation of the optical system to be characterised.} \textbf{a}, Response of the system without the presence of a quantum emitter. An input laser gets transmitted (reflected) by the photonic structure with coefficient $|t_0|^{2}$ ($|r_0|^{2}$). \textbf{b}, With the presence of a single emitter, the laser interferes with the resonance fluorescence of the driven quantum system, producing Fano transmission and reflection spectra. \textbf{c}, Example Markovian structures covered by our general characterisation model including a continuous waveguide, a resonator in the weak coupling regime, and an interrupted waveguide.}
	\label{fig:Extinction setup}
\end{figure}

\noindent \textbf{Theoretical Framework. }We consider the general system depicted in \figRef{\ref{fig:Extinction setup}(a)}, in which two optical guiding structures, labelled as the pump and probe waveguides, are connected by a photonic black box. We are interested in the probability $\beta_{\text{probe}}$ (or $\beta_{\text{pump}}$) that a photon leaving the emitter will be coupled into the probe (or pump) waveguide. In order to measure these probabilities, optical power $P_{\text{in}}$ is coupled into transverse mode $m$ of the pump guide, making a field ${\operatorname{Re}\{ \mathcal{E}_{\text{in}}\mathbf{u}_{m}(x,y)e^{i (k z - \omega t)}\} }$. Here, $\mathbf{u}_{m}(x,y)$ gives the transverse distribution of the field in mode $m$, $z$ is the direction of propagation and $\mathcal{E}_{\text{in}}$ is the amplitude of the pump light in that mode. In the absence of the emitter, the transmitted pump light in mode $m$ of the probe guide is ${\operatorname{Re}\{t_{0} \mathcal{E}_{\text{in}} \mathbf{u}_{m}(x,y)e^{i(k z - \omega t)}\} }$, where $t_{0}$ is the complex transmission factor.

Consider an emitter with an upper level $|e\rangle$ and a lower level $|g\rangle$ (other levels are sufficiently far from resonance that they can be adiabatically eliminated from the coherent dynamics). When the emitter is put in place, the field has the option of being scattered by the emitter into the probe waveguide, as depicted in \figRef{\ref{fig:Extinction setup}(b)}. We make two key assumptions about the dynamics. First, we assume that the optical reservoir decays faster than all other relevant time scales \cite{Dung2002}. This Markovian approximation is justified for most single-photon sources where a fast optical response is desirable. Second, we make the semi-classical assumption that the quantum correlations between the pump field and the emitter can be ignored \cite{Allen1987}. With these assumptions, the total output field in the probe waveguide is given by \cite{Asenjo-Garcia2017} \begin{equation}
    \label{eq:efield}
    \mathcal{E}_{\text{out}}(\mathbf{r}) =t_{0}\mathcal{E}_{\text{in}} + \mathcal{E}^{t}_{\text{emit}} \sigma^{-}   \ ,
\end{equation}

\noindent where we have dropped the factor $\mathbf{u}_{m}(x,y)e^{i(k z - \omega t)}$ from both sides of the equation. The operator ${\sigma^{-}=\ket{g}\bra{e}}e^{i \omega t}$ ensures that the emission of a photon is accompanied by de-excitation of the emitter. Let the total power scattered by the emitter at the frequency $\omega$ be $P_{\text{emit}}$, a fraction $\beta_{\text{probe}}$ of which is scattered into the probe guide mode $m$. Since the power in the guide is proportional to the square of the field it follows that

\begin{equation} \label{eq:probepower}
    \frac{|\mathcal{E}^{t}_{\text{emit}}|^{2} \langle \sigma^{+}\sigma^{-} \rangle}{|\mathcal{E_{\text{in}}}|^{2}} = \frac{\beta_{\text{probe}} P_{\text{emit}}}{P_{\text{in}}} \,,
\end{equation}

\noindent where $\sigma^{+}=e^{-i \omega t}\ket{e}\bra{g}$ and the angle brackets indicate the steady-state expectation value of the atomic operator. With continuous-wave pumping in the near-resonant regime, \begin{equation} \label{eq:Pemit}
P_{\text{emit}}=\hbar\omega\gamma_{1}\braket{\sigma^{+}\sigma^{-}}\,,
\end{equation}

\noindent where $\gamma_{1}$ is the population decay rate of the excited state due to radiation at the frequency $\omega$ of the pump light. This may be a partial decay rate because Raman sidebands and any non-radiative decay processes are not included here. In order to evaluate this, we need to know the field that drives the emitter. In the \hyperref[Methods]{Methods} section we show that this is related to $\beta_{\text{pump}}$ through the relation \begin{equation} \label{eq:OmegaSquared}
   \Omega^{2} = 4\beta_{\text{pump}} \gamma_{\text{1}} \frac{P_{\text{in}}}{\hbar \omega} \,.
\end{equation}

\noindent Here $\Omega$ is the Rabi frequency, defined as ${\mathbf{d}\cdot\boldsymbol{\mathbf{E}}(\mathbf{r}_{0}) / \hbar}$, where $\mathbf{d}$ is the dipole transition matrix element and the pump field at the site of the emitter is ${\operatorname{Re}\{\mathbf{E}(\mathbf{r}_{0})  e^{-i \omega t}\} }$. We choose $\Omega$ to be real without loss of generality.

On substituting  \eqRef{\ref{eq:Pemit}} and \eqRef{\ref{eq:OmegaSquared}} into \eqRef{\ref{eq:probepower}}, we find

\begin{equation}
    |\mathcal{E}^{t}_{\text{emit}}| = \sqrt{4 \beta_{\text{pump}}\beta_{\text{probe}}} \frac{\gamma_{1}}{\Omega} |\mathcal{E_{\text{in}}}| \ .
\end{equation}

\noindent Hence, ignoring a global phase, the field at the output end of the guide is given by

\begin{equation}
    \mathcal{E}_{\text{out}} = \left(|t_{0}| +  \frac{\beta_{\text{eff}}\gamma_{1}}{\Omega} e^{i\phi_{\text{T}}}\sigma^{-} \right) |\mathcal{E}_{\text{in}}| \ .
\end{equation}

\noindent Here we have introduced $\phi_{\text{T}}$, which is the phase difference between the two transmitted fields due to propagation; a further phase shift will come from the lag of the dipole response $\sigma^{-}$. We have also replaced $\sqrt{4\beta_{\text{pump}}\beta_{\text{probe}}}$ by $\beta_{\text{eff}}$. Note that $\beta_{\text{pump}}$ and $\beta_{\text{probe}}$ are both between 0 and 1 but $\beta_{\text{pump}}\beta_{\text{probe}} \leq \beta_{\text{pump}}(1-\beta_{\text{pump}}) \leq 1/4$, so the best case is $\beta_{\text{eff}} = 1$. It follows that the net transmission power is given by

\begin{align}
    \label{eq:outpow}
	\begin{aligned}
        &\frac{P_{\text{out}}}{P_{\text{in}}} = \frac{\langle \mathcal{E}_{\text{out}} \mathcal{E}^{\dagger}_{\text{out}} \rangle }{|\mathcal{E}_{\text{in}}|^2} \\ &= |t_{0}|^{2} + 2|t_{0}| \frac{\beta_{\text{eff}}\gamma_{1}}{\Omega} \text{Re}\left( e^{-i\phi_{\text{T}}}\rho_{ge} \right) + \left( \frac{\beta_{\text{eff}}\gamma_{1}}{\Omega} \right) ^{2} \rho_{ee} \ ,
	\end{aligned}
\end{align}

\noindent where $\rho$ is the density matrix of the emitter with  $\rho_{ge} = \langle \sigma^{+} \rangle$, and $\rho_{ee} = \langle \sigma^{+}\sigma^{-}\rangle$. These three terms correspond respectively to the transmitted pump power, the interference term between the pump field and the coherently scattered field, and the scattered power, all in the probe guide.

The density matrix elements are found by solving the optical Bloch equations \cite{Asenjo-Garcia2017, Turschmann2019}, with the result \begin{align}
\label{eq:rho}
    \begin{aligned}
    \rho_{ee} &=  \frac{\frac{1}{2}S}{(\delta\omega / \Gamma_{2})^{2} + 1 + S} \ , \\
	\rho_{ge} &= -  \frac{\Omega / (2\Gamma_{2})}{(\delta\omega / \Gamma_{2})^{2} + 1 + S} (\frac{\delta\omega}{\Gamma_{2}} + i) \ ,
	\end{aligned}
\end{align}

\noindent where $\delta\omega = \omega - \omega_{0}$ is the detuning of the laser from resonance and $S = \frac{\Omega^{2}}{\Gamma_{1}\Gamma_{2}}$ is the saturation parameter. Here $\Gamma_{1}$ is the total decay rate of the upper state population, while $\Gamma_{2}$ is the decay rate of the coherence $\rho_{ge}$ by all decoherence mechanisms. On substituting \eqRef{\ref{eq:rho}} into \eqRef{\ref{eq:outpow}} we obtain the transmission spectrum

\begin{align}
	\begin{aligned}
       \frac{P_{\text{out}}}{P_{\text{in}}} =  |t_{0}|^2 - \bigg\{ 2\alpha \beta_{\text{eff}}|t_{0}| \left( \sin(\phi_{\text{T}}) + \frac{\delta\omega}{\Gamma_{2}} \cos(\phi_{\text{T}})  \right) \\ - (\alpha \beta_{\text{eff}})^2 \bigg\} \frac{\Gamma_{1}/(2 \Gamma_{2})}{(\delta\omega / \Gamma_{2})^{2} + 1 + S} \ ,
	\end{aligned}
	\label{eq:transmission}
\end{align}

\noindent where $\alpha=\gamma_{1}/\Gamma_{1}$. A similar analysis gives the reflection spectrum

\begin{align}
    \begin{aligned}
        \frac{P_{\text{refl}}}{P_{\text{in}}} = |r_{0}|^2 - \bigg\{ 4\alpha \beta_{\text{pump}}|r_{0}| \left( \sin(\phi_{\text{R}}) + \frac{\delta\omega}{\Gamma_{2}} \cos(\phi_{\text{R}})  \right) \\ - 4(\alpha \beta_{\text{pump}})^2 \bigg\} \frac{\Gamma_{1}/(2 \Gamma_{2})}{(\delta\omega / \Gamma_{2})^{2} + 1 + S} \,,
	\end{aligned}
	\label{eq:reflection}
\end{align}
	
\noindent where $r_{0}$ is the reflection coefficient and $\phi_{\text{R}}$ is the reflection analogue of $\phi_{\text{T}}$.

In an experiment to measure the transmission as a function of frequency, the spectrum may be fitted to \eqRef{\ref{eq:transmission}}. When $S \ll 1$ and the value of $\Gamma_{1}/(2\Gamma_{2})$ is known, the fit will yield values for $|t_{0}|$, $\alpha\beta_{\text{eff}}$ and $\phi_{\text{T}}$. However, it is common in a real experiment for the light to be attenuated by the train of auxiliary optics so that the measured powers $\mathcal{P_{\text{out}}}$ and $\mathcal{P_{\text{in}}}$ have the ratio ${\mathcal{P_{\text{out}}} /\mathcal{P_{\text{in}}} = \eta P_{\text{out}}/P_{\text{in}}}$, and the value of $\eta$ is unknown. For large detuning, the measured transmission $\mathcal{P}_{\text{out}}/\mathcal{P}_{\text{in}}$ then takes the value $\eta|t_{0}|^2$. On normalising the data to this transmission we have from \eqRef{\ref{eq:transmission}}

\begin{align}
	\begin{aligned}
        T = \frac{\mathcal{P}_{\text{out}}}{\eta|t_{0}|^2\mathcal{P}_{\text{in}}} =  1 - \frac{\alpha \beta_{\text{eff}}}{|t_{0}|} \bigg\{ 2 \left( \sin(\phi_{\text{T}}) + \frac{\delta\omega}{\Gamma_{2}} \cos(\phi_{\text{T}})  \right) \\ - \frac{ \alpha \beta_{\text{eff}}}{|t_{0}|} \bigg\} \frac{\Gamma_{1}/(2 \Gamma_{2})}{(\delta\omega / \Gamma_{2})^{2} + 1 + S} \ . 	\end{aligned}
 	\label{eq:transmission2}
\end{align}

\noindent In this case, the fit yields a value for $\alpha\beta_{\text{eff}}/|t_{0}|$, rather than $\alpha\beta_{\text{eff}}$. One may determine $|t_{0}|$ by an auxiliary experiment which compares the device with another that contains no emitter and has $|t_{0}|=1$. Alternatively, for structures where a normal mode decomposition is appropriate, for example a continuous waveguide or weak cavity as depicted in \figRef{\ref{fig:Extinction setup}(c)}, most of the parameters in \eqRef{\ref{eq:transmission}} and \eqRef{\ref{eq:reflection}} can be calculated analytically, as we consider further in the Supplementary Information.\\
\\


\begin{figure*}
	\begin{center}
	\includegraphics{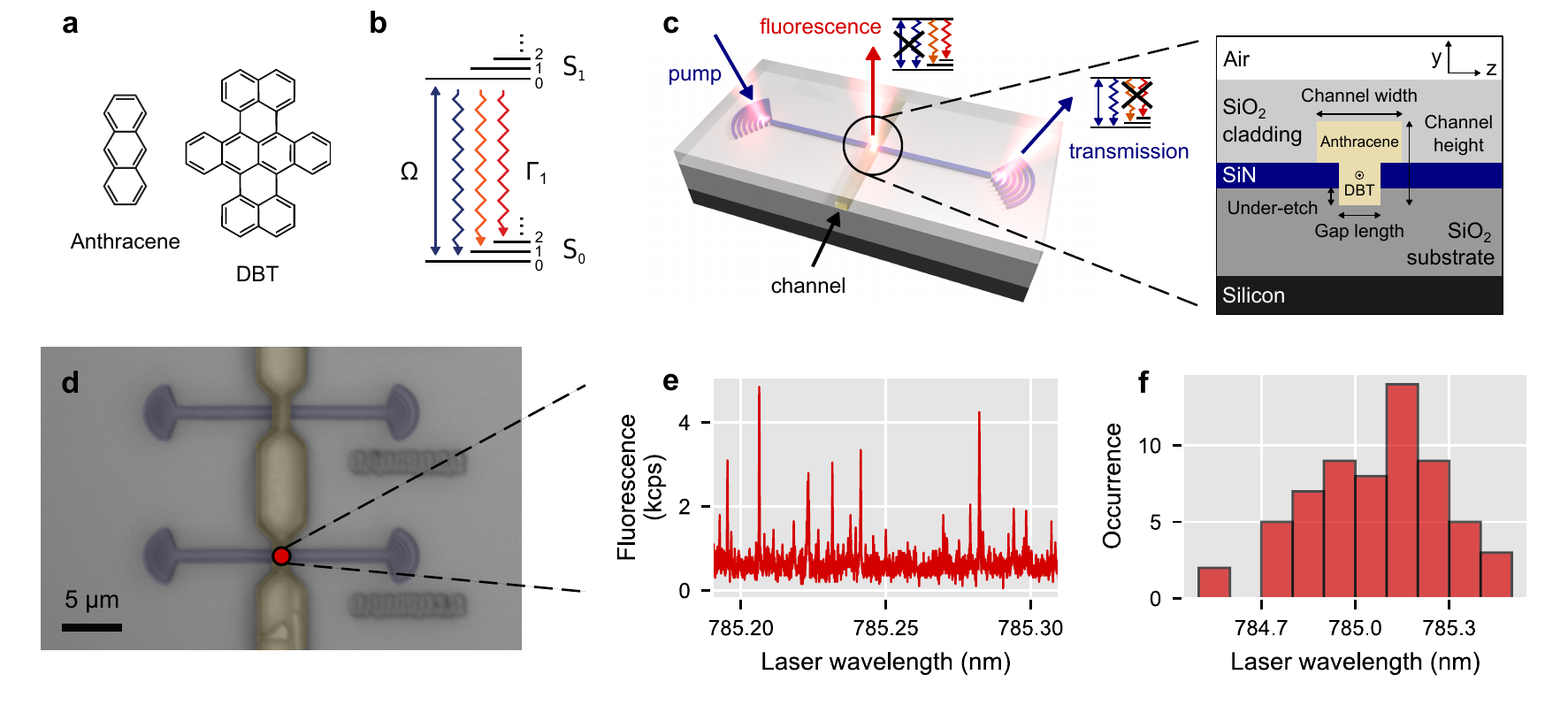}
	\end{center}
	\vspace{-5mm}
	\caption{\textbf{Localised growth of DBT-doped anthracene in the vicinity of an interrupted waveguide.} \textbf{a}, Molecular structure of anthracene and DBT . \textbf{b}, Jablonski diagram of relevant DBT levels in anthracene: triplets are ignored because inter-system crossing is very weak \cite{Nicolet2007a}. \textbf{c}, Overview of the grating couplers, the interrupted waveguide and the  microfluidic channel crossing it. The zoom-in shows details of the intersection between guide and channel. \textbf{d}, False-color optical-microscope image of two devices with the microfluidic channels filled. \textbf{e}, Fluorescence excitation spectrum of molecules near the gap in a device at cryogenic temperature. The molecules are excited from the `pump' waveguide and fluorescence is collected by the confocal microscope from the red dot shown in \textbf{d}. \textbf{f}, Wavelength distribution of the DBT resonances from the same confocal spot. }
	\label{fig:Geometry}	
\end{figure*}

\noindent \textbf{Micro-fluidic integration of single molecules with waveguides. }In the early days of single-emitter spectroscopy, it was found that large polycyclic aromatic hydrocarbon (PAH) molecules such as pentacene \cite{Moerner1989}, terrylene \cite{Nicolet2006}, dibenzanthanthrene (DBATT) \cite{Jelezko1997} and dibenzoterrylene (DBT) \cite{Nicolet2007a} could be hosted in PAH crystals to form stable quantum-emitters in the solid state. In this work we use  DBT-doped anthracene. The molecular structures are shown in \figRef{\ref{fig:Geometry}(a)} and relevant energy levels of DBT are drawn in \figRef{\ref{fig:Geometry}(b)}. This well-studied combination has a very weak singlet-triplet inter-system crossing, is highly photostable, has a high probability of radiative decay on the zero-phonon line (ZPL) (shown blue) \cite{Nicolet2007a, Nicolet2007b}, and has a lifetime-limited resonance width at cryogenic temperatures \cite{Trebbia2009}.

The coupling of photons to single PAH molecules has been used in bulk material to demonstrate, for example, a single-molecule optical transistor \cite{Hwang2009} and few-photon nonlinear optics \cite{Maser2016}, but for applications such as a deterministic photon source stronger coupling is desirable. A natural way to achieve that is to integrate the emitters into a photonic structure \cite{Hwang2011a,Turschmann2017, Lombardi2018, Rattenbacher2019}, and it is convenient to grow doped organic crystals around the structure by solidifying from a molten mixture \cite{Faez2014, Turschmann2017}. Normally, the structure is made of inorganic material and the organic molecule couples to an evanescent field. However, the molecule is usually unstable at less than a few hundred nanometres from the inorganic/organic interface \cite{Gmeiner2016a}, and therefore it can only be placed in the tail of the evanescent field where the dipolar coupling to the photonic mode is weak.  Here, we take a different approach, shown in \figRef{\ref{fig:Geometry}(c)}, where a silicon nitride waveguide having grating couplers at each end is interrupted by a sub-wavelength gap. After fabricating the waveguide chip, molten anthracene doped with DBT is drawn by capillary forces along a microfluidic channel which cuts across the waveguide and fills the gap, as depicted in \figRef{\ref{fig:Geometry}(c)}. A numerical simulation, details of which are given in the Supplementary Information, shows that the coupling efficiency $\beta_{\text{eff}}$ for a molecule sitting at the center of the gap decreases rapidly with the length of the gap. However, with a gap of \SI{400}{\nano\meter} this can be as high as \SI{30}{\%}. A smaller gap can yield higher coupling, but the guide faces on each end of the gap are then close enough to the molecule that they may compromise its optical properties. The coupling can, of course, be much higher with the introduction of a cavity  \cite{Rattenbacher2019}.

We began device fabrication with a silicon wafer that had a layer of thermal oxide covered by silicon nitride, and we  patterned the interrupted waveguides and grating couplers in the silicon nitride. The micro-fluidic channels were then fabricated from a sacrificial resist layer on top of which SiO\textsubscript{2} was sputtered. We cleaved the chips to expose the channel entrance at the facets and baked the sample to remove the resist, thereby opening hollow channels. Finally, we filled the channels with DBT-doped crystalline anthracene by controlled heating and subsequent cooling. See \hyperref[Methods]{Methods} for details of device fabrication and filling of the capillaries.

\begin{figure*}
	\begin{center}
	\includegraphics{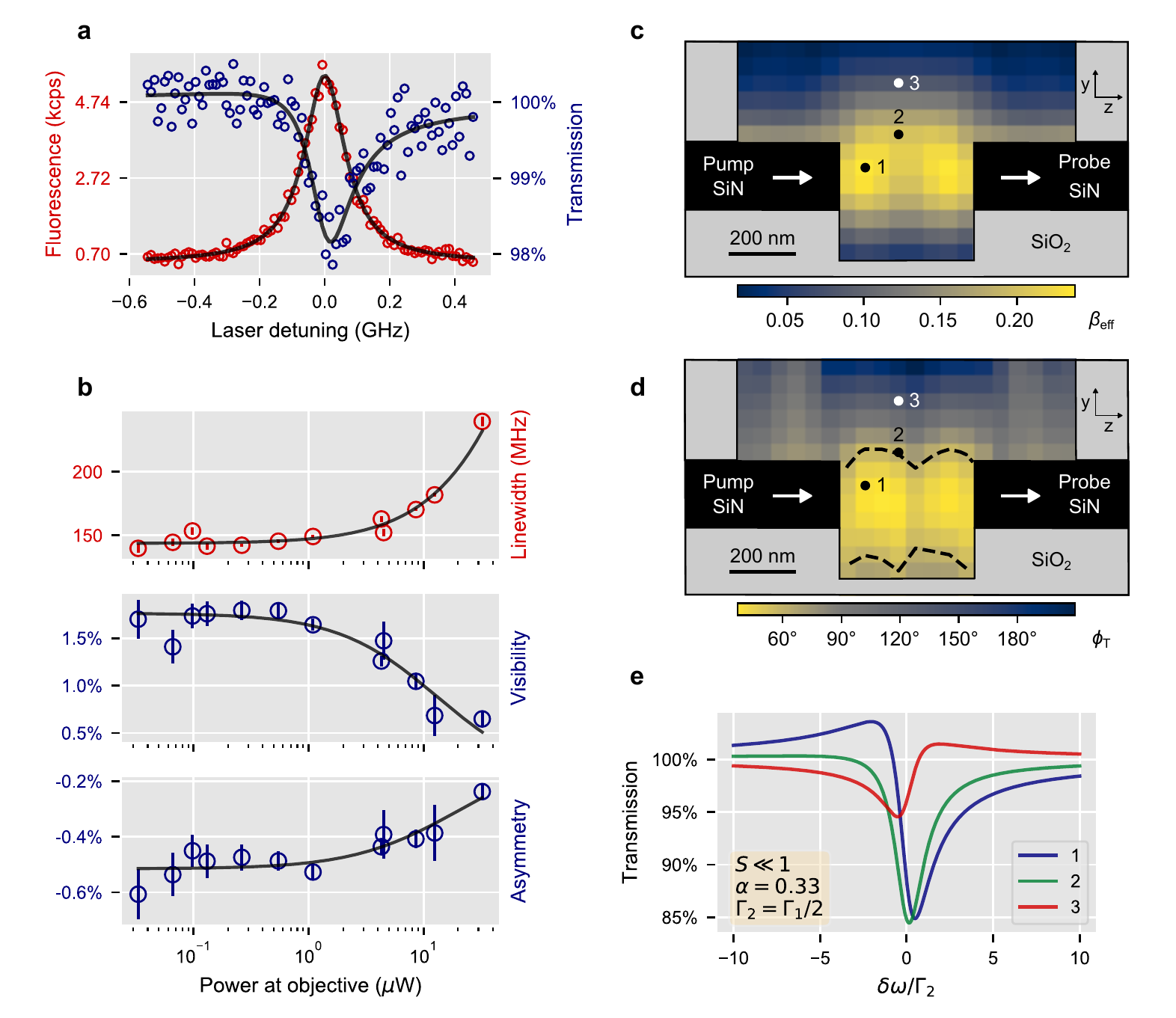}
	\end{center}
	\vspace{-3mm}
	\caption{\textbf{Coherent characterisation of a coupled single molecule by extinction spectroscopy.} \textbf{a}, Experimental data. Red circles: red-shifted fluorescence emerging from the gap, fitted with a Lorentzian curve. Blue circles: \SI{785}{\nano\meter} transmission observed through the output grating, fitted using \eqRef{\ref{eq:transmission2}}. \textbf{b}, Plots of linewidth (FWHM), visibility ($V$) and asymmetry ($q$) as a function of the pump power. \textbf{c}, \textbf{d}, FDTD computation of the coupling efficiency $\beta_{\text{eff}}$ and propagation phase difference $\phi_{\text{T}}$ for a dipole along the $x$-direction, placed in the mirror symmetry plane of the structure. \textbf{e}, Expected transmission spectrum for lifetime-limited DBT molecules placed at various positions inside the channel.}
	\label{fig:ExpResults}
\end{figure*}

To verify that we had stable emitters in the vicinity of the waveguide, we performed fluorescence spectroscopy at cryogenic temperature under a microscope (see \hyperref[Methods]{Methods} for a detailed description of the optical setup). A waveguide chip was filled with DBT-doped anthracene at $10^{-4}$ molar fraction, then cooled in the cryostat to \SI{4.7}{\kelvin} and positioned so that a device having \SI{400}{\nano\meter} gap length and \SI{1}{\micro\meter} channel width (see \figRef{\ref{fig:Geometry}(c)}) was in focus at the centre of the field of view. \figRef{\ref{fig:Geometry}(d)} shows a false-colour white light image of the structure. A cw laser was focused onto one of the grating couplers to excite molecules from the `pump' side of the waveguide, and was continuously scanned at low power between \SIrange{784.5}{785.5}{\nano\meter} to cover the inhomogeneous width of the $S_{0,0} \leftrightarrow S_{1,0}$ transition. Light was then collected from the vicinity of the waveguide gap and sent to a photon counter.  An \SI{800}{\nano\meter} long-pass filter removed any scattered laser light together with the ZPL fluorescence, leaving only the red-shifted fluorescence. We plot a slice of the scan in \figRef{\ref{fig:Geometry}(e)} which reveals the characteristic Lorentzian resonance peaks of many DBT molecules having a range of resonant frequencies. A histogram of these frequencies is given in \figRef{\ref{fig:Geometry}(f)}. We estimate the concentration of molecules resonating at \SI{785}{\nano\meter} in the gap to be $\approx \SI{160}{molecules\per\micro\meter\cubed\per\nano\meter}$. Some of the light is collected from molecules that are well away from the gap but are excited by scattered pump light, and these have poor coupling to the guide. To characterise the strength of the coupling, we therefore use the extinction spectroscopy method developed above, as described next.\\ 


\noindent \textbf{Characterisation of the coupling.} We collected light from the grating on the `probe side' of the waveguide and used a $785\pm3$~nm band-pass filter to remove the red-shifted fluorescence and most of the local phonon sideband \cite{Clear2020}. (We show in the Supplementary Information that imperfect filtering has a negligible effect). The grating selectively couples to the $x$-polarised waveguide mode, which is the mode that we pump. We scanned the pump frequency over the resonance of a single DBT molecule and recorded both the red-shifted fluorescence from the gap and the resonant transmission from the output grating, as indicated in \figRef{\ref{fig:Geometry}(c)}. The data are plotted in \figRef{\ref{fig:ExpResults}(a)} as open circles in red and blue for the fluorescence and transmission respectively. A solid line shows the least-squares fit of a Lorentzian to the fluorescence data, which gives us the linewidth, $2\Gamma_2\sqrt{1+S}$ (full width at half maximum). Knowing that, we then fit \eqRef{\ref{eq:transmission2}} to the transmission data to produce the solid line through the blue data points. For this second fit we  express \eqRef{\ref{eq:transmission2}} as the Fano lineshape \begin{equation}
    \label{fano}
    T(\epsilon) = \frac{1 - (V + q^{2})}{1 + \epsilon^{2}} + \frac{\left(q+\epsilon\right)^{2}}{1+\epsilon^{2}}\,,
\end{equation}

\noindent with \begin{gather} \label{eq:visibility}
    V = \beta \left( 2 \sin(\phi_{\text{T}}) - \beta \right)  \frac{\Gamma_{1}/(2 \Gamma_{2})}{1+S}\,, \\  \label{eq:assymery}
    q=-\beta\cos(\phi_{\text{T}})\frac{\Gamma_{1}/(2 \Gamma_{2})}{\sqrt{1+S}}\,,
\end{gather}

\noindent where $\epsilon=\delta\omega/(\Gamma_{2}\sqrt{1+S})$ is the normalised detuning and $\beta=\alpha \beta_{\text{eff}}/|t_{0}|$ is the scaled coupling efficiency. This fit gives the values of $V$ and $q$. Repeated scans at twelve different pump powers gave us values for the FWHM linewidths, visibilities ($V$) and asymmetries ($q$) that are plotted in \figRef{\ref{fig:ExpResults}(b)}. On extrapolating to the limit of low power, we find the values $\Gamma_{2} / \pi = \SI{144(2)}{\mega\hertz}$, $V_{0} = \SI{1.8(1)}{\%}$ and $q_{0} = \SI{-5.2(1)e-3}{}$. The lifetime of the $S_1$ state is $\Gamma_1^{-1}=(4.5\pm1)\,$ns \cite{Grandi2016a}, giving a minimum linewidth of $\sim35\,$MHz and $\Gamma_{1}/(2\Gamma_{2}) = 0.25$. That is significantly less than 1 because our cryostat only cooled the sample down to $\sim \SI{4.7}{\kelvin}$, whereas the minimum width is reached at $\sim3.5\,$K. \eqRef{\ref{eq:visibility}} and \eqRef{\ref{eq:assymery}} give two solutions for $\beta$ and $\phi_{\text{T}}$. In the limit of small $S$,

\begin{gather} \label{eq:betasol}
    \beta_{\pm} = \sqrt{2-\tilde{V}_{0} \pm 2 \sqrt{1-\tilde{q}_{0}^{2} - \tilde{V}_{0}}}\,, \\ \label{eq:phisol} \begin{aligned}
    \phi_{T\pm} = \operatorname{atan2}\bigg( 4 \tilde{q}_{0}^{2} - & \tilde{V}_{0} ( \tilde{V}_{0} +  \beta_{\pm}^2 - 4)~, \\ & 2 \tilde{q}_{0} (2 \tilde{V}_{0} + \beta_{\pm}^2 - 4) \bigg) \,,
    \end{aligned}
\end{gather}

\noindent where $\tilde{q}_{0} = q_{0}/(\Gamma_{1}/(2\Gamma_{2}))$ and $\tilde{V}_{0} = V_{0}/(\Gamma_{1}/(2\Gamma_{2}))$. We use the function $\operatorname{atan2}(\text{numerator, denominator})$ to ensure that $\phi_{\text{T}}$ is placed in the correct quadrant.

In order to derive $\beta_{\text{eff}}$ from $V_0$ and $q_0$, we measured $|t_{0}|$ by comparing the off-resonant transmission of the device with the transmission of a second device, which was identical except that the waveguide had no gap. We scanned the laser frequency to look for possible cavity resonances in the optical setup, which would have invalidated the method, but found only a very weak modulation. This comparison gave $|t_{0}|=0.63(6)$, which differs slightly from the numerically calculated transmission $|t_{\text{sim}}| = 0.81$, perhaps because our simulation simplifies the anisotropic refractive index of the anthracene. Having measured $|t_{0}|$ and setting $\alpha=0.33$ (known from bulk measurements of DBT in anthracene \cite{Trebbia2009, Clear2020}), we find that the $\beta_{+}$ solution gives the unphysical result $\beta_{\text{eff}}>1$, so we conclude that $\beta_{\text{eff}} = \beta_{-} |t_{0}|/\alpha =9(2)\%$, with the error bar coming roughly equally from the uncertainty in $|t_{0}|$ and from the other uncertainties combined. The corresponding solution for the phase difference is $\phi_{\text{T}} = \SI{61(2)}{\degree}$ (independent of $|t_{0}|$), with the error bar coming primarily from the uncertainties in $q_0$ and $V_0$.

It is instructive to compare these results for $\beta_{\text{eff}}$ and $\phi_{\text{T}}$ with a numerical simulation (see \hyperref[Methods]{Methods}). \figRef{\ref{fig:ExpResults}(c)} shows $\beta_{\text{eff}}$ for a dipole transverse to the guide (along $x$), placed in the $yz$ plane centred on the guide. (The coupling at the centre of the \SI{400}{\nano\meter} gap is less than the maximum possible $30\%$ because the height and width of the guide are not perfectly optimised). The coupling is strongest for an emitter placed in the gap, but we note that an emitter outside the gap and close to the guide couples to the evanescent field, as seen by the yellow strip running along the outside of the guide. \figRef{\ref{fig:ExpResults}(d)} shows the propagation phase difference $\phi_{\text{T}}$. This phase varies strongly with position in the gap, in contrast to the behaviour when coupling to a cavity. Also, we find that $\phi_{\text{T}}$ tends to 90{\degree} when the emitter couples to the evanescent field on the side of of the guide and far from the gap, as expected for coupling to a continuous waveguide. See the Supplementary Information for treatments of the continuous waveguide and weak cavity cases. In \figRef{\ref{fig:ExpResults}(e)} we plot the transmission spectra calculated for weakly-pumped, ideally-polarised DBT molecules at each of the three positions marked in \figRef{\ref{fig:ExpResults}(c, d)}.

The dashed lines in \figRef{\ref{fig:ExpResults}(d)} show where a dipole lying in the $yz$ plane through the centre of the guide would give the measured value $\phi_{\text{T}} = \SI{61}{\degree}$. If the molecule is in this plane, we expect it to be near the upper contour, for example in the position marked $2$, because the lower one is too close to the substrate for photo-stability. On this line the  calculated coupling efficiency varies in the range $20-21\%$, which is to be compared with the $9(2)\%$ we have measured. Our molecule has no reason to be aligned along $x$, so the simulation would be consistent with our measurement if the molecule makes an angle of $\theta=49\degree$ to the $x$-axis. Of course, there is also no reason for the molecule to sit in the plane $x=0$. Looking at the whole surface where $\phi_{\text{T}} = \SI{61}{\degree}$, we find that the simulated coupling varies in the range $11-21\%$, and conclude therefore that $\theta$ is in the range $25\degree-49\degree$.

\section*{Discussion}

We have demonstrated how to characterise the coherent scattering of light by a single quantum emitter, in a photonic environment that cannot be decomposed into a small number of relevant modes. We have shown that the transmission and reflection spectra are described by Fano lineshapes, from which one can extract the coupling efficiencies without needing precise knowledge of the photonic structure. Our method generalises extinction spectroscopy to complex geometries, yielding values for coupling efficiency without needing to measure in detail all the losses in the system. Further, the propagation phase shift $\phi_{\text{T}}$ can provide some information on the position of the emitter within the structure and on the orientation of its transition dipole.

We have also demonstrated a new way to integrate a single molecule into photonic structures on a chip by using microfluidic channels to bring doped crystals to the desired locations. We hope that increased control over the microfluidic channel geometry will allow fabrication of more complex structures to boost the collection efficiency. Specifically, slotted waveguides \cite{Hwang2011a} and slotted photonic crystal waveguides \cite{DiFalco2008a} are promising ways to achieve this. This work also opens the possibility of integrating molecular quantum emitters with photonic components such as beam splitters, interferometers and detectors, to study quantum networks and integrated quantum sensors \cite{Muschik2014}. In addition, we have shown that anthracene crystals can be highly doped to achieve densities on a chip of hundreds of emitters per $\lambda^{3}$ per nm. This could enable the study of collective behaviour of coupled quantum systems such as polaritonic light–matter states \cite{Turschmann2019} or direct dipole-dipole interactions \cite{Hettich2002}.

	
\section*{Methods} \label{Methods}

\noindent \textbf{Derivation of Equation 4.}  The classical pump field $\{\mathbf{E}^{f}_{\text{m}},\mathbf{H}^{f}_{\text{m}}\}$ propagates forward (toward the black box) in transverse mode $m$ at frequency $\omega$ and with power $P_{\text{in}}$.  This field leaves the guide and enters the black box, where it induces a dipole moment $\mathbf{D}$ in an emitter that radiates  the field $\{\mathbf{E}_{\mathbf{d}},\mathbf{H}_{\mathbf{d}}\}$ with power $P_{\text{d}}$. A fraction $\beta_{\text{pump}}$ of that radiated power goes back into the pump guide. From the orthogonality of modes \cite{Oskooi2013} we have

\begin{equation}
\beta_{\text{pump}} = \frac{|\frac{1}{4}\int \left( \mathbf{E}_{\mathbf{d}} \times \left(\mathbf{H}^{b}_{\text{m}} \right)^* + \left( \mathbf{E}^{b}_{\text{m}} \right)^* \times \mathbf{H}_{\mathbf{d}} \right) \cdot \mathrm{d} \mathbf{S}|^2}{P_{\text{in}} \ P_{\mathbf{d}}}\,,  \\ \label{eq:mode expansion}
\end{equation}

\noindent\noindent where the superscript $b$ denotes the mode propagating backwards (away from the black box).

Wanting to relate these fields to $\mathbf{D}$, we note that the dipole at position $\mathbf{r}_{0}$ produces a current density $\mathbf{j}_{\mathbf{d}}(\mathbf{r}) = - i \omega \mathbf{D} \delta(\mathbf{r} - \mathbf{r}_{0})$. Similarly, the pump field may be viewed as the result of (fictitious) electric and magnetic current densities $\mathbf{j}_{\text{in}}(\mathbf{r}) = \mathbf{\delta n} \times \mathbf{H}^{f}_{\text{m}}$ and $\mathbf{m}_{\text{in}}(\mathbf{r}) = \mathbf{\delta n} \times \mathbf{E}^{f}_{\text{m}}$ \cite{Oskooi2013}. These lie on a plane surface $S$ far from the black box, whose normal is parallel to the direction of propagation, and $\mathbf{\delta n}$ is a Dirac delta function along the normal. Now we can make use of the reciprocity theorem \cite{Novotny2009} to write

\begin{equation}
	\int \mathbf{j}_{\mathbf{d}} \cdot \mathbf{E} \ \mathrm{d} V = \int \left(\mathbf{j}_{\text{in}} \cdot \mathbf{E}_{\mathbf{d}} + \mathbf{m}_{\text{in}} \cdot \mathbf{H}_{\mathbf{d}}\right) \ \mathrm{d} V \,,
\end{equation}
\noindent where $\mathbf{E}$ is the pump field and the integrals are over an arbitrarily large volume that includes $S$. On evaluating these integrals with the explicit current densities we find that

\begin{equation}
  -i\,\omega\,\mathbf{D}\cdot \mathbf{E}(\mathbf{r}_{0})=\int \left( \mathbf{E}_{\mathbf{d}} \times \left(\mathbf{H}^{b}_{\text{m}} \right)^* + \left( \mathbf{E}^{b}_{\text{m}} \right)^* \times \mathbf{H}_{\mathbf{d}} \right) \cdot \mathrm{d} \mathbf{S}\,, \\ \label{eq:reciprocity}
\end{equation}

\noindent where we have used the relations ${\mathbf{E}^{f}_{\text{m}} = \left(\mathbf{E}^{b}_{\text{m}} \right)^*}$ and ${\mathbf{H}^{f}_{\text{m}} = - \left(\mathbf{H}^{b}_{\text{m}} \right)^*}$. Using \eqRef{\ref{eq:reciprocity}} to eliminate the integral in \eqRef{\ref{eq:mode expansion}} we have

\begin{equation}
	\beta_{\text{pump}}	= \frac{1}{16} \frac{\omega^{2} |\mathbf{D}\cdot \mathbf{E}(\mathbf{r}_{0})|^2 }{P_{\text{in}} \ P_{\mathbf{d}}} \,. \\ \label{eq:almost there}
\end{equation}

\noindent Connecting the classical dipole to the quantum emitter, we replace the ratio $|\mathbf{D}\cdot\mathbf{E}(\mathbf{r}_{0})|^2/P_{\text{d}}$ by $(2\hbar\Omega)^2/(\hbar\omega\gamma_1)$ \cite{Novotny2009}. Both $\gamma_1$ and the Rabi frequency $\Omega$ are defined in the main text. With this substitution in \eqRef{\ref{eq:almost there}}  we obtain the result given in \eqRef{\ref{eq:OmegaSquared}}.
\vspace{3mm}

\noindent \textbf{Device fabrication. }The waveguides are fabricated from a \SI{200}{\nano\meter} thick silicon nitride layer on \SI{2}{\micro\meter} of silica on silicon. The waveguide patterns are first written into ma-N 2403 resist by electron beam lithography and transferred into the underlying Si\textsubscript{3}N\textsubscript{4} layer by reactive ion etching with a CHF\textsubscript{3} plasma. We over-etch the silicon nitride by \SI{150}{\nano\meter} so that the middle of the waveguide sits \SI{250}{\nano\meter} away from the bottom surface. In this way, the position of maximum coupling is not too close to the bottom surface. The waveguides on the chip have a width of \SI{400}{\nano\meter} and gap lengths ranging from no gap to \SI{400}{\nano\meter}. We terminate the waveguides with gratings based on concentric circles. To avoid reflections, the gratings are designed to couple light at an angle of  \SI{10}{\degree} to the vertical.

To overlay the micro-fluidic channels, we first spin-coat a \SI{1}{\micro\meter} layer of AZ nLOF 2020 resist which is diluted 4:1 (resist:solvent w/w) with PGMEA. Electron beam lithography exposes the resist along channels that are perpendicular to the waveguides and aligned with the gaps. We then deposit \SI{2}{\micro\meter} of SiO\textsubscript{2} on top of the resist using RF-sputtering. Next, the sample is cleaved to expose the resist channels on both facets. Finally, we place the sample in a furnace which is heated to \SI{550}{\celsius} in ambient atmosphere. Under these conditions, we find that the resist is released from the channels without leaving any residue, and we are left with open structures which can be filled with molten DBT-doped anthracene.
\vspace{3mm}

\noindent \textbf{Capillary filling. }In order to fill the micro-fluidic channels with doped-anthracene, we use growth from the melt by solidification \cite{Faez2014, Turschmann2017}. We first place a small quantity of DBT-doped anthracene powder (\SI{e-4}{mol/mol} concentration) on the facets of the chip. The sample is then put on a hotplate in a glove box which is continuously purged with nitrogen. We heat the sample at a rate of \SI{5}{\celsius\per\second} and hold the temperature at \SI{210}{\celsius} until the channels are visibly filled by the melted material. Finally, we cool the sample at a rate of \SI{-5}{\celsius\per\second} causing the anthracene to crystallise. This yields long stretches of the capillaries filled by solid anthracene. We check the quality of the DBT molecules in the capillaries using cryogenic fluorescence spectroscopy and we show in the Supplementary Information that the spectral stability is not appreciably affected by the constrained geometry of the micro-fluidic channel.
\vspace{3mm}

\noindent \textbf{Optical setup. }The optical apparatus was a three-beam confocal microscope built around a closed-cycle cryostat (Cryostation, Montana Instruments), as illustrated in Fig. S2 of the Supplementary Information.  The primary excitation light came from a continuously tunable titanium:sapphire laser (SolsTiS, MSquared) that was power-stabilised using an acousto-optic modulator and a proportional-integrated-derivative controller (SIM960, SRS). The light was delivered to the apparatus through a single-mode fibre, then collimated with an aspheric lens and polarised before passing through a half-wave plate and a bandpass filter (F1) to produce a linearly-polarised beam with adjustable polarisation angle and spectral purity. This entered a 10\% transmission (90\% reflection) beamsplitter (BS), and the transmitted light was sent to a pair of electronically controlled galvanometer mirrors (GM).  Through the use of two lenses in a `4f' configuration (L1, L2), the angular change in the galvanometer mirrors allowed us to adjust the angle of incidence onto an objective lens (LD EC Epiplan-Neofluor 100x, 0.75NA, Zeiss) inside the cryostat without translating across the objective aperture. This in turn caused a focused spot to be raster scanned across the sample. The back aperture of the objective was overfilled to ensure the minimum spot size of 720\,nm full-width half-maximum. The sample was mounted on a 3-axis piezo-controlled translation stage (PS, Attocube) which we used to locate waveguides and bring them into focus. Molecule fluorescence followed the beam path back to the 90:10 BS where the 90\% reflected portion passed through a long-pass filter (F2) to remove the excitation laser before being collected in multimode fiber and detected on a silicon avalanche photodiode. By inserting a pellicle BS into the excitation path after the scanning mirrors we introduced white light (WL) from a lamp onto the sample. This light was then reflected from the sample and off another pellicle BS above the cryostat to an electron multiplying charge coupled device (CCD) camera (iXon, Andor) which took wide-field images, such as that shown in \figRef{\ref{fig:Geometry}(d)}. A second single-mode fibre input (shown within the rectangle labelled ``Grating Coupling'') was collimated, polarised, filtered and steered onto a (90:10) beam splitter, before being combined with the main beam path in the `4f' lens setup using a 50:50 beam splitter. The steering mirrors allowed the beam to couple into the pump guide through its grating coupler, giving a typical total coupling efficiency of $8\%$ from fibre to waveguide. Light emerging from the probe guide grating coupler was directed back to a final single mode fibre (in the rectangle) and thence to the detector that recorded the transmission spectrum.
\vspace{3mm}

\noindent \textbf{Finite-difference time-domain simulations. }The numerical simulations of the device are performed with three-dimensional finite-difference time-domain (FDTD) analysis using the Meep software package \cite{Oskooi2010}. The structural parameters, as defined in \figRef{\ref{fig:Geometry}(c)}, are: waveguide width~=~\SI{400}{\nano\meter}, waveguide height~=~\SI{200}{\nano\meter}, gap length~=~\SI{400}{\nano\meter}, under-etch~=~\SI{150}{\nano\meter}, channel width~=~\SI{1}{\micro\meter} and channel height~=~\SI{1}{\micro\meter}. We use a mesh size of \SI{16}{\nano\meter} and perfectly matched layers to simulate open boundaries. Anthracene is a biaxial material but for simplicity we choose to approximate it as isotropic with refractive index $n=1.8$.

To compute the transmission through the gap, we use a continuous eigensource to excite the $x$-polarised mode of the pump waveguide. We determine the power transmitted into the $x$-polarised mode of the probe waveguide by projecting the field at the output end onto that mode. For coupling efficiency calculations, we use a continuous dipole source placed at a given position in the channel and monitor the total power emitted together with the power coupled into the $x$-polarised modes of the waveguides.

FDTD simulations also allow us to calculate the phase difference $\phi_{\text{T}}$. Using a continuous eigensource to excite the $x$-polarised mode $m$ of the pump waveguide, we first compute the phase shift of the transmitted light, $\operatorname{Arg}(t_0)$, which is the phase difference between light in mode $m$ at the entrance of the probe guide and the exit of the pump guide. Mode decomposition is used to isolated the field coupled to mode $m$ of the probe waveguide. For the propagation phase shift of the scattered light, we place an electric dipole at the position of the molecule. The dipole oscillates in phase with the pump field at that position, but the pump field is not turned on. Again we take the difference between the phase of the (dipole) field in mode $m$ at the entrance to the probe guide and that of the pump field (if it were turned on) at the exit of the pump guide. Calling this latter phase shift $\Delta\phi$, we have $\phi_{\text{T}}=\Delta\phi-\operatorname{Arg}(t_0)$.
\vspace{3mm}



\section*{Acknowledgments}
\noindent We thank Jon Dyne and Dave Pitman for their expert mechanical workshop support. We also thank David Mack and Javier Cambiasso for their help with nanofabrication. This work was supported by EPSRC (EP/P030130/1, EP/P01058X/1, EP/R044031/1, EP/P510257/1, and EP/L016524/1), the Royal Society (UF160475, RGF/R1/180066, and RGF/EA/180203), and the EraNET Cofund Initiative QuantERA under the European Union’s Horizon 2020 research and innovation programme, Grant No. 731473 (ORQUID Project).
\vspace{3mm}

\section*{Author Contributions}
\noindent S.B. and E.A.H. formulated the theory; S.B. performed the numerical simulations; S.B., S.N., L.J., A.O. and W.H.P.P. designed and fabricated the nanophotonic devices; S.B., R.C.S., K.D.M. and A.S.C. built the experiment and took the data; S.B., R.C.S., K.D.M., E.A.H. and A.S.C. analysed the data. All authors discussed the results. S.B. and E.A.H. wrote the initial draft manuscript, and all authors contributed to the final manuscript. E.A.H. and A.S.C. led the project.


\begin{thebibliography}{51}%
\makeatletter
\providecommand \@ifxundefined [1]{%
 \@ifx{#1\undefined}
}%
\providecommand \@ifnum [1]{%
 \ifnum #1\expandafter \@firstoftwo
 \else \expandafter \@secondoftwo
 \fi
}%
\providecommand \@ifx [1]{%
 \ifx #1\expandafter \@firstoftwo
 \else \expandafter \@secondoftwo
 \fi
}%
\providecommand \natexlab [1]{#1}%
\providecommand \enquote  [1]{``#1''}%
\providecommand \bibnamefont  [1]{#1}%
\providecommand \bibfnamefont [1]{#1}%
\providecommand \citenamefont [1]{#1}%
\providecommand \href@noop [0]{\@secondoftwo}%
\providecommand \href [0]{\begingroup \@sanitize@url \@href}%
\providecommand \@href[1]{\@@startlink{#1}\@@href}%
\providecommand \@@href[1]{\endgroup#1\@@endlink}%
\providecommand \@sanitize@url [0]{\catcode `\\12\catcode `\$12\catcode
  `\&12\catcode `\#12\catcode `\^12\catcode `\_12\catcode `\%12\relax}%
\providecommand \@@startlink[1]{}%
\providecommand \@@endlink[0]{}%
\providecommand \url  [0]{\begingroup\@sanitize@url \@url }%
\providecommand \@url [1]{\endgroup\@href {#1}{\urlprefix }}%
\providecommand \urlprefix  [0]{URL }%
\providecommand \Eprint [0]{\href }%
\providecommand \doibase [0]{https://doi.org/}%
\providecommand \selectlanguage [0]{\@gobble}%
\providecommand \bibinfo  [0]{\@secondoftwo}%
\providecommand \bibfield  [0]{\@secondoftwo}%
\providecommand \translation [1]{[#1]}%
\providecommand \BibitemOpen [0]{}%
\providecommand \bibitemStop [0]{}%
\providecommand \bibitemNoStop [0]{.\EOS\space}%
\providecommand \EOS [0]{\spacefactor3000\relax}%
\providecommand \BibitemShut  [1]{\csname bibitem#1\endcsname}%
\let\auto@bib@innerbib\@empty
\bibitem [{\citenamefont {Crespi}\ \emph {et~al.}(2012)\citenamefont {Crespi},
  \citenamefont {Lobino}, \citenamefont {Matthews}, \citenamefont {Politi},
  \citenamefont {Neal}, \citenamefont {Ramponi}, \citenamefont {Osellame},\
  and\ \citenamefont {O'Brien}}]{Crespi2012}%
  \BibitemOpen
  \bibfield  {author} {\bibinfo {author} {\bibfnamefont {A.}~\bibnamefont
  {Crespi}}, \bibinfo {author} {\bibfnamefont {M.}~\bibnamefont {Lobino}},
  \bibinfo {author} {\bibfnamefont {J.~C.}\ \bibnamefont {Matthews}}, \bibinfo
  {author} {\bibfnamefont {A.}~\bibnamefont {Politi}}, \bibinfo {author}
  {\bibfnamefont {C.~R.}\ \bibnamefont {Neal}}, \bibinfo {author}
  {\bibfnamefont {R.}~\bibnamefont {Ramponi}}, \bibinfo {author} {\bibfnamefont
  {R.}~\bibnamefont {Osellame}},\ and\ \bibinfo {author} {\bibfnamefont
  {J.~L.}\ \bibnamefont {O'Brien}},\ }\bibfield  {title} {\bibinfo {title}
  {{Measuring protein concentration with entangled photons}},\ }\href
  {https://doi.org/10.1063/1.4724105} {\bibfield  {journal} {\bibinfo
  {journal} {Applied Physics Letters}\ }\textbf {\bibinfo {volume} {100}},\
  \bibinfo {pages} {233704} (\bibinfo {year} {2012})},\ \Eprint
  {https://arxiv.org/abs/1109.3128} {arXiv:1109.3128} \BibitemShut {NoStop}%
\bibitem [{\citenamefont {Ono}\ \emph {et~al.}(2019)\citenamefont {Ono},
  \citenamefont {Sinclair}, \citenamefont {Bonneau}, \citenamefont {Thompson},
  \citenamefont {Matthews},\ and\ \citenamefont {Rarity}}]{Ono2019}%
  \BibitemOpen
  \bibfield  {author} {\bibinfo {author} {\bibfnamefont {T.}~\bibnamefont
  {Ono}}, \bibinfo {author} {\bibfnamefont {G.~F.}\ \bibnamefont {Sinclair}},
  \bibinfo {author} {\bibfnamefont {D.}~\bibnamefont {Bonneau}}, \bibinfo
  {author} {\bibfnamefont {M.~G.}\ \bibnamefont {Thompson}}, \bibinfo {author}
  {\bibfnamefont {J.~C.~F.}\ \bibnamefont {Matthews}},\ and\ \bibinfo {author}
  {\bibfnamefont {J.~G.}\ \bibnamefont {Rarity}},\ }\bibfield  {title}
  {\bibinfo {title} {{Observation of nonlinear interference on a silicon
  photonic chip}},\ }\href {https://doi.org/10.1364/ol.44.001277} {\bibfield
  {journal} {\bibinfo  {journal} {Optics Letters}\ }\textbf {\bibinfo {volume}
  {44}},\ \bibinfo {pages} {1277} (\bibinfo {year} {2019})}\BibitemShut
  {NoStop}%
\bibitem [{\citenamefont {Sparrow}\ \emph {et~al.}(2018)\citenamefont
  {Sparrow}, \citenamefont {Mart{\'{i}}n-L{\'{o}}pez}, \citenamefont
  {Maraviglia}, \citenamefont {Neville}, \citenamefont {Harrold}, \citenamefont
  {Carolan}, \citenamefont {Joglekar}, \citenamefont {Hashimoto}, \citenamefont
  {Matsuda}, \citenamefont {O'Brien}, \citenamefont {Tew},\ and\ \citenamefont
  {Laing}}]{Sparrow2018}%
  \BibitemOpen
  \bibfield  {author} {\bibinfo {author} {\bibfnamefont {C.}~\bibnamefont
  {Sparrow}}, \bibinfo {author} {\bibfnamefont {E.}~\bibnamefont
  {Mart{\'{i}}n-L{\'{o}}pez}}, \bibinfo {author} {\bibfnamefont
  {N.}~\bibnamefont {Maraviglia}}, \bibinfo {author} {\bibfnamefont
  {A.}~\bibnamefont {Neville}}, \bibinfo {author} {\bibfnamefont
  {C.}~\bibnamefont {Harrold}}, \bibinfo {author} {\bibfnamefont
  {J.}~\bibnamefont {Carolan}}, \bibinfo {author} {\bibfnamefont {Y.~N.}\
  \bibnamefont {Joglekar}}, \bibinfo {author} {\bibfnamefont {T.}~\bibnamefont
  {Hashimoto}}, \bibinfo {author} {\bibfnamefont {N.}~\bibnamefont {Matsuda}},
  \bibinfo {author} {\bibfnamefont {J.~L.}\ \bibnamefont {O'Brien}}, \bibinfo
  {author} {\bibfnamefont {D.~P.}\ \bibnamefont {Tew}},\ and\ \bibinfo {author}
  {\bibfnamefont {A.}~\bibnamefont {Laing}},\ }\bibfield  {title} {\bibinfo
  {title} {{Simulating the vibrational quantum dynamics of molecules using
  photonics}},\ }\href {https://doi.org/10.1038/s41586-018-0152-9} {\bibfield
  {journal} {\bibinfo  {journal} {Nature}\ }\textbf {\bibinfo {volume} {557}},\
  \bibinfo {pages} {660} (\bibinfo {year} {2018})}\BibitemShut {NoStop}%
\bibitem [{\citenamefont {Qiang}\ \emph {et~al.}(2018)\citenamefont {Qiang},
  \citenamefont {Zhou}, \citenamefont {Wang}, \citenamefont {Wilkes},
  \citenamefont {Loke}, \citenamefont {O'Gara}, \citenamefont {Kling},
  \citenamefont {Marshall}, \citenamefont {Santagati}, \citenamefont {Ralph},
  \citenamefont {Wang}, \citenamefont {O'Brien}, \citenamefont {Thompson},\
  and\ \citenamefont {Matthews}}]{Qiang2018}%
  \BibitemOpen
  \bibfield  {author} {\bibinfo {author} {\bibfnamefont {X.}~\bibnamefont
  {Qiang}}, \bibinfo {author} {\bibfnamefont {X.}~\bibnamefont {Zhou}},
  \bibinfo {author} {\bibfnamefont {J.}~\bibnamefont {Wang}}, \bibinfo {author}
  {\bibfnamefont {C.~M.}\ \bibnamefont {Wilkes}}, \bibinfo {author}
  {\bibfnamefont {T.}~\bibnamefont {Loke}}, \bibinfo {author} {\bibfnamefont
  {S.}~\bibnamefont {O'Gara}}, \bibinfo {author} {\bibfnamefont
  {L.}~\bibnamefont {Kling}}, \bibinfo {author} {\bibfnamefont {G.~D.}\
  \bibnamefont {Marshall}}, \bibinfo {author} {\bibfnamefont {R.}~\bibnamefont
  {Santagati}}, \bibinfo {author} {\bibfnamefont {T.~C.}\ \bibnamefont
  {Ralph}}, \bibinfo {author} {\bibfnamefont {J.~B.}\ \bibnamefont {Wang}},
  \bibinfo {author} {\bibfnamefont {J.~L.}\ \bibnamefont {O'Brien}}, \bibinfo
  {author} {\bibfnamefont {M.~G.}\ \bibnamefont {Thompson}},\ and\ \bibinfo
  {author} {\bibfnamefont {J.~C.}\ \bibnamefont {Matthews}},\ }\bibfield
  {title} {\bibinfo {title} {{Large-scale silicon quantum photonics
  implementing arbitrary two-qubit processing}},\ }\href
  {https://doi.org/10.1038/s41566-018-0236-y} {\bibfield  {journal} {\bibinfo
  {journal} {Nature Photonics}\ }\textbf {\bibinfo {volume} {12}},\ \bibinfo
  {pages} {534} (\bibinfo {year} {2018})},\ \Eprint
  {https://arxiv.org/abs/1809.09791} {arXiv:1809.09791} \BibitemShut {NoStop}%
\bibitem [{\citenamefont {Lodahl}\ \emph {et~al.}(2015)\citenamefont {Lodahl},
  \citenamefont {Mahmoodian},\ and\ \citenamefont {Stobbe}}]{Lodahl2015}%
  \BibitemOpen
  \bibfield  {author} {\bibinfo {author} {\bibfnamefont {P.}~\bibnamefont
  {Lodahl}}, \bibinfo {author} {\bibfnamefont {S.}~\bibnamefont {Mahmoodian}},\
  and\ \bibinfo {author} {\bibfnamefont {S.}~\bibnamefont {Stobbe}},\
  }\bibfield  {title} {\bibinfo {title} {{Interfacing single photons and single
  quantum dots with photonic nanostructures}},\ }\href
  {https://doi.org/10.1103/RevModPhys.87.347} {\bibfield  {journal} {\bibinfo
  {journal} {Reviews of Modern Physics}\ }\textbf {\bibinfo {volume} {87}},\
  \bibinfo {pages} {347} (\bibinfo {year} {2015})},\ \Eprint
  {https://arxiv.org/abs/1312.1079} {arXiv:1312.1079} \BibitemShut {NoStop}%
\bibitem [{\citenamefont {Aharonovich}\ \emph {et~al.}(2016)\citenamefont
  {Aharonovich}, \citenamefont {Englund},\ and\ \citenamefont
  {Toth}}]{Aharonovich2016}%
  \BibitemOpen
  \bibfield  {author} {\bibinfo {author} {\bibfnamefont {I.}~\bibnamefont
  {Aharonovich}}, \bibinfo {author} {\bibfnamefont {D.}~\bibnamefont
  {Englund}},\ and\ \bibinfo {author} {\bibfnamefont {M.}~\bibnamefont
  {Toth}},\ }\bibfield  {title} {\bibinfo {title} {{Solid-state single-photon
  emitters}},\ }\href {https://doi.org/10.1038/nphoton.2016.186} {\bibfield
  {journal} {\bibinfo  {journal} {Nature Photonics}\ }\textbf {\bibinfo
  {volume} {10}},\ \bibinfo {pages} {631} (\bibinfo {year} {2016})}\BibitemShut
  {NoStop}%
\bibitem [{\citenamefont {Lounis}\ and\ \citenamefont
  {Moerner}(2000)}]{Lounis2000}%
  \BibitemOpen
  \bibfield  {author} {\bibinfo {author} {\bibfnamefont {B.}~\bibnamefont
  {Lounis}}\ and\ \bibinfo {author} {\bibfnamefont {W.~E.}\ \bibnamefont
  {Moerner}},\ }\bibfield  {title} {\bibinfo {title} {{Single photons on demand
  from a single molecule at room temperature}},\ }\href
  {https://doi.org/10.1038/35035032} {\bibfield  {journal} {\bibinfo  {journal}
  {Nature}\ }\textbf {\bibinfo {volume} {407}},\ \bibinfo {pages} {491}
  (\bibinfo {year} {2000})}\BibitemShut {NoStop}%
\bibitem [{\citenamefont {Hwang}\ and\ \citenamefont
  {Hinds}(2011)}]{Hwang2011a}%
  \BibitemOpen
  \bibfield  {author} {\bibinfo {author} {\bibfnamefont {J.}~\bibnamefont
  {Hwang}}\ and\ \bibinfo {author} {\bibfnamefont {E.~A.}\ \bibnamefont
  {Hinds}},\ }\bibfield  {title} {\bibinfo {title} {{Dye molecules as
  single-photon sources and large optical nonlinearities on a chip}},\ }\href
  {https://doi.org/10.1088/1367-2630/13/8/085009} {\bibfield  {journal}
  {\bibinfo  {journal} {New Journal of Physics}\ }\textbf {\bibinfo {volume}
  {13}},\ \bibinfo {pages} {085009} (\bibinfo {year} {2011})}\BibitemShut
  {NoStop}%
\bibitem [{\citenamefont {Javadi}\ \emph {et~al.}(2015)\citenamefont {Javadi},
  \citenamefont {S{\"{o}}llner}, \citenamefont {Arcari}, \citenamefont
  {Hansen}, \citenamefont {Midolo}, \citenamefont {Mahmoodian}, \citenamefont
  {Kir{\v{s}}anskė}, \citenamefont {Pregnolato}, \citenamefont {Lee},
  \citenamefont {Song}, \citenamefont {Stobbe},\ and\ \citenamefont
  {Lodahl}}]{Javadi2015}%
  \BibitemOpen
  \bibfield  {author} {\bibinfo {author} {\bibfnamefont {A.}~\bibnamefont
  {Javadi}}, \bibinfo {author} {\bibfnamefont {I.}~\bibnamefont
  {S{\"{o}}llner}}, \bibinfo {author} {\bibfnamefont {M.}~\bibnamefont
  {Arcari}}, \bibinfo {author} {\bibfnamefont {S.~L.}\ \bibnamefont {Hansen}},
  \bibinfo {author} {\bibfnamefont {L.}~\bibnamefont {Midolo}}, \bibinfo
  {author} {\bibfnamefont {S.}~\bibnamefont {Mahmoodian}}, \bibinfo {author}
  {\bibfnamefont {G.}~\bibnamefont {Kir{\v{s}}anskė}}, \bibinfo {author}
  {\bibfnamefont {T.}~\bibnamefont {Pregnolato}}, \bibinfo {author}
  {\bibfnamefont {E.~H.}\ \bibnamefont {Lee}}, \bibinfo {author} {\bibfnamefont
  {J.~D.}\ \bibnamefont {Song}}, \bibinfo {author} {\bibfnamefont
  {S.}~\bibnamefont {Stobbe}},\ and\ \bibinfo {author} {\bibfnamefont
  {P.}~\bibnamefont {Lodahl}},\ }\bibfield  {title} {\bibinfo {title}
  {{Single-photon nonlinear optics with a quantum dot in a waveguide}},\ }\href
  {https://doi.org/10.1038/ncomms9655} {\bibfield  {journal} {\bibinfo
  {journal} {Nature Communications}\ }\textbf {\bibinfo {volume} {6}},\
  \bibinfo {pages} {8655} (\bibinfo {year} {2015})},\ \Eprint
  {https://arxiv.org/abs/1504.06895} {arXiv:1504.06895} \BibitemShut {NoStop}%
\bibitem [{\citenamefont {Wang}\ \emph
  {et~al.}(2019{\natexlab{a}})\citenamefont {Wang}, \citenamefont {He},
  \citenamefont {Chung}, \citenamefont {Hu}, \citenamefont {Yu}, \citenamefont
  {Chen}, \citenamefont {Ding}, \citenamefont {Chen}, \citenamefont {Qin},
  \citenamefont {Yang}, \citenamefont {Liu}, \citenamefont {Duan},
  \citenamefont {Li}, \citenamefont {Gerhardt}, \citenamefont {Winkler},
  \citenamefont {Jurkat}, \citenamefont {Wang}, \citenamefont {Gregersen},
  \citenamefont {Huo}, \citenamefont {Dai}, \citenamefont {Yu}, \citenamefont
  {H{\"{o}}fling}, \citenamefont {Lu},\ and\ \citenamefont {Pan}}]{Wang2019}%
  \BibitemOpen
  \bibfield  {author} {\bibinfo {author} {\bibfnamefont {H.}~\bibnamefont
  {Wang}}, \bibinfo {author} {\bibfnamefont {Y.~M.}\ \bibnamefont {He}},
  \bibinfo {author} {\bibfnamefont {T.~H.}\ \bibnamefont {Chung}}, \bibinfo
  {author} {\bibfnamefont {H.}~\bibnamefont {Hu}}, \bibinfo {author}
  {\bibfnamefont {Y.}~\bibnamefont {Yu}}, \bibinfo {author} {\bibfnamefont
  {S.}~\bibnamefont {Chen}}, \bibinfo {author} {\bibfnamefont {X.}~\bibnamefont
  {Ding}}, \bibinfo {author} {\bibfnamefont {M.~C.}\ \bibnamefont {Chen}},
  \bibinfo {author} {\bibfnamefont {J.}~\bibnamefont {Qin}}, \bibinfo {author}
  {\bibfnamefont {X.}~\bibnamefont {Yang}}, \bibinfo {author} {\bibfnamefont
  {R.~Z.}\ \bibnamefont {Liu}}, \bibinfo {author} {\bibfnamefont {Z.~C.}\
  \bibnamefont {Duan}}, \bibinfo {author} {\bibfnamefont {J.~P.}\ \bibnamefont
  {Li}}, \bibinfo {author} {\bibfnamefont {S.}~\bibnamefont {Gerhardt}},
  \bibinfo {author} {\bibfnamefont {K.}~\bibnamefont {Winkler}}, \bibinfo
  {author} {\bibfnamefont {J.}~\bibnamefont {Jurkat}}, \bibinfo {author}
  {\bibfnamefont {L.~J.}\ \bibnamefont {Wang}}, \bibinfo {author}
  {\bibfnamefont {N.}~\bibnamefont {Gregersen}}, \bibinfo {author}
  {\bibfnamefont {Y.~H.}\ \bibnamefont {Huo}}, \bibinfo {author} {\bibfnamefont
  {Q.}~\bibnamefont {Dai}}, \bibinfo {author} {\bibfnamefont {S.}~\bibnamefont
  {Yu}}, \bibinfo {author} {\bibfnamefont {S.}~\bibnamefont {H{\"{o}}fling}},
  \bibinfo {author} {\bibfnamefont {C.~Y.}\ \bibnamefont {Lu}},\ and\ \bibinfo
  {author} {\bibfnamefont {J.~W.}\ \bibnamefont {Pan}},\ }\bibfield  {title}
  {\bibinfo {title} {{Towards optimal single-photon sources from polarized
  microcavities}},\ }\href {https://doi.org/10.1038/s41566-019-0494-3}
  {\bibfield  {journal} {\bibinfo  {journal} {Nature Photonics}\ }\textbf
  {\bibinfo {volume} {13}},\ \bibinfo {pages} {770} (\bibinfo {year}
  {2019}{\natexlab{a}})}\BibitemShut {NoStop}%
\bibitem [{\citenamefont {Arcari}\ \emph {et~al.}(2014)\citenamefont {Arcari},
  \citenamefont {S{\"{o}}llner}, \citenamefont {Javadi}, \citenamefont
  {{Lindskov Hansen}}, \citenamefont {Mahmoodian}, \citenamefont {Liu},
  \citenamefont {Thyrrestrup}, \citenamefont {Lee}, \citenamefont {Song},
  \citenamefont {Stobbe},\ and\ \citenamefont {Lodahl}}]{Arcari2014}%
  \BibitemOpen
  \bibfield  {author} {\bibinfo {author} {\bibfnamefont {M.}~\bibnamefont
  {Arcari}}, \bibinfo {author} {\bibfnamefont {I.}~\bibnamefont
  {S{\"{o}}llner}}, \bibinfo {author} {\bibfnamefont {A.}~\bibnamefont
  {Javadi}}, \bibinfo {author} {\bibfnamefont {S.}~\bibnamefont {{Lindskov
  Hansen}}}, \bibinfo {author} {\bibfnamefont {S.}~\bibnamefont {Mahmoodian}},
  \bibinfo {author} {\bibfnamefont {J.}~\bibnamefont {Liu}}, \bibinfo {author}
  {\bibfnamefont {H.}~\bibnamefont {Thyrrestrup}}, \bibinfo {author}
  {\bibfnamefont {E.~H.}\ \bibnamefont {Lee}}, \bibinfo {author} {\bibfnamefont
  {J.~D.}\ \bibnamefont {Song}}, \bibinfo {author} {\bibfnamefont
  {S.}~\bibnamefont {Stobbe}},\ and\ \bibinfo {author} {\bibfnamefont
  {P.}~\bibnamefont {Lodahl}},\ }\bibfield  {title} {\bibinfo {title}
  {{Near-Unity Coupling Efficiency of a Quantum Emitter to a Photonic Crystal
  Waveguide}},\ }\href {https://doi.org/10.1103/PhysRevLett.113.093603}
  {\bibfield  {journal} {\bibinfo  {journal} {Physical Review Letters}\
  }\textbf {\bibinfo {volume} {113}},\ \bibinfo {pages} {093603} (\bibinfo
  {year} {2014})},\ \Eprint {https://arxiv.org/abs/1402.2081} {arXiv:1402.2081}
  \BibitemShut {NoStop}%
\bibitem [{\citenamefont {T{\"{u}}rschmann}\ \emph {et~al.}(2019)\citenamefont
  {T{\"{u}}rschmann}, \citenamefont {{Le Jeannic}}, \citenamefont {Simonsen},
  \citenamefont {Haakh}, \citenamefont {G{\"{o}}tzinger}, \citenamefont
  {Sandoghdar}, \citenamefont {Lodahl},\ and\ \citenamefont
  {Rotenberg}}]{Turschmann2019}%
  \BibitemOpen
  \bibfield  {author} {\bibinfo {author} {\bibfnamefont {P.}~\bibnamefont
  {T{\"{u}}rschmann}}, \bibinfo {author} {\bibfnamefont {H.}~\bibnamefont {{Le
  Jeannic}}}, \bibinfo {author} {\bibfnamefont {S.~F.}\ \bibnamefont
  {Simonsen}}, \bibinfo {author} {\bibfnamefont {H.~R.}\ \bibnamefont {Haakh}},
  \bibinfo {author} {\bibfnamefont {S.}~\bibnamefont {G{\"{o}}tzinger}},
  \bibinfo {author} {\bibfnamefont {V.}~\bibnamefont {Sandoghdar}}, \bibinfo
  {author} {\bibfnamefont {P.}~\bibnamefont {Lodahl}},\ and\ \bibinfo {author}
  {\bibfnamefont {N.}~\bibnamefont {Rotenberg}},\ }\bibfield  {title} {\bibinfo
  {title} {{Coherent nonlinear optics of quantum emitters in nanophotonic
  waveguides}},\ }\href {https://doi.org/10.1515/nanoph-2019-0126} {\bibfield
  {journal} {\bibinfo  {journal} {Nanophotonics}\ }\textbf {\bibinfo {volume}
  {8}},\ \bibinfo {pages} {1641} (\bibinfo {year} {2019})},\ \Eprint
  {https://arxiv.org/abs/1906.08565} {arXiv:1906.08565} \BibitemShut {NoStop}%
\bibitem [{\citenamefont {Gerhardt}\ \emph {et~al.}(2007)\citenamefont
  {Gerhardt}, \citenamefont {Wrigge}, \citenamefont {Bushev}, \citenamefont
  {Zumofen}, \citenamefont {Agio}, \citenamefont {Pfab},\ and\ \citenamefont
  {Sandoghdar}}]{Gerhardt2007a}%
  \BibitemOpen
  \bibfield  {author} {\bibinfo {author} {\bibfnamefont {I.}~\bibnamefont
  {Gerhardt}}, \bibinfo {author} {\bibfnamefont {G.}~\bibnamefont {Wrigge}},
  \bibinfo {author} {\bibfnamefont {P.}~\bibnamefont {Bushev}}, \bibinfo
  {author} {\bibfnamefont {G.}~\bibnamefont {Zumofen}}, \bibinfo {author}
  {\bibfnamefont {M.}~\bibnamefont {Agio}}, \bibinfo {author} {\bibfnamefont
  {R.}~\bibnamefont {Pfab}},\ and\ \bibinfo {author} {\bibfnamefont
  {V.}~\bibnamefont {Sandoghdar}},\ }\bibfield  {title} {\bibinfo {title}
  {{Strong extinction of a laser beam by a single molecule}},\ }\href
  {https://doi.org/10.1103/PhysRevLett.98.033601} {\bibfield  {journal}
  {\bibinfo  {journal} {Physical Review Letters}\ }\textbf {\bibinfo {volume}
  {98}},\ \bibinfo {pages} {033601} (\bibinfo {year} {2007})}\BibitemShut
  {NoStop}%
\bibitem [{\citenamefont {Pototschnig}\ \emph {et~al.}(2011)\citenamefont
  {Pototschnig}, \citenamefont {Chassagneux}, \citenamefont {Hwang},
  \citenamefont {Zumofen}, \citenamefont {Renn},\ and\ \citenamefont
  {Sandoghdar}}]{Pototschnig2011}%
  \BibitemOpen
  \bibfield  {author} {\bibinfo {author} {\bibfnamefont {M.}~\bibnamefont
  {Pototschnig}}, \bibinfo {author} {\bibfnamefont {Y.}~\bibnamefont
  {Chassagneux}}, \bibinfo {author} {\bibfnamefont {J.}~\bibnamefont {Hwang}},
  \bibinfo {author} {\bibfnamefont {G.}~\bibnamefont {Zumofen}}, \bibinfo
  {author} {\bibfnamefont {A.}~\bibnamefont {Renn}},\ and\ \bibinfo {author}
  {\bibfnamefont {V.}~\bibnamefont {Sandoghdar}},\ }\bibfield  {title}
  {\bibinfo {title} {{Controlling the phase of a light beam with a single
  molecule}},\ }\href {https://doi.org/10.1103/PhysRevLett.107.063001}
  {\bibfield  {journal} {\bibinfo  {journal} {Physical Review Letters}\
  }\textbf {\bibinfo {volume} {107}},\ \bibinfo {pages} {063001} (\bibinfo
  {year} {2011})},\ \Eprint {https://arxiv.org/abs/1103.6048} {arXiv:1103.6048}
  \BibitemShut {NoStop}%
\bibitem [{\citenamefont {Foster}\ \emph {et~al.}(2019)\citenamefont {Foster},
  \citenamefont {Hallett}, \citenamefont {Iorsh}, \citenamefont {Sheldon},
  \citenamefont {Godsland}, \citenamefont {Royall}, \citenamefont {Clarke},
  \citenamefont {Shelykh}, \citenamefont {Fox}, \citenamefont {Skolnick},
  \citenamefont {Itskevich},\ and\ \citenamefont {Wilson}}]{Foster2019}%
  \BibitemOpen
  \bibfield  {author} {\bibinfo {author} {\bibfnamefont {A.~P.}\ \bibnamefont
  {Foster}}, \bibinfo {author} {\bibfnamefont {D.}~\bibnamefont {Hallett}},
  \bibinfo {author} {\bibfnamefont {I.~V.}\ \bibnamefont {Iorsh}}, \bibinfo
  {author} {\bibfnamefont {S.~J.}\ \bibnamefont {Sheldon}}, \bibinfo {author}
  {\bibfnamefont {M.~R.}\ \bibnamefont {Godsland}}, \bibinfo {author}
  {\bibfnamefont {B.}~\bibnamefont {Royall}}, \bibinfo {author} {\bibfnamefont
  {E.}~\bibnamefont {Clarke}}, \bibinfo {author} {\bibfnamefont {I.~A.}\
  \bibnamefont {Shelykh}}, \bibinfo {author} {\bibfnamefont {A.~M.}\
  \bibnamefont {Fox}}, \bibinfo {author} {\bibfnamefont {M.~S.}\ \bibnamefont
  {Skolnick}}, \bibinfo {author} {\bibfnamefont {I.~E.}\ \bibnamefont
  {Itskevich}},\ and\ \bibinfo {author} {\bibfnamefont {L.~R.}\ \bibnamefont
  {Wilson}},\ }\bibfield  {title} {\bibinfo {title} {{Tunable Photon Statistics
  Exploiting the Fano Effect in a Waveguide}},\ }\href
  {https://doi.org/10.1103/PhysRevLett.122.173603} {\bibfield  {journal}
  {\bibinfo  {journal} {Physical Review Letters}\ }\textbf {\bibinfo {volume}
  {122}},\ \bibinfo {pages} {173603} (\bibinfo {year} {2019})},\ \Eprint
  {https://arxiv.org/abs/1811.08860} {arXiv:1811.08860} \BibitemShut {NoStop}%
\bibitem [{\citenamefont {Hwang}\ \emph {et~al.}(2009)\citenamefont {Hwang},
  \citenamefont {Pototschnig}, \citenamefont {Lettow}, \citenamefont {Zumofen},
  \citenamefont {Renn}, \citenamefont {G{\"{o}}tzinger},\ and\ \citenamefont
  {Sandoghdar}}]{Hwang2009}%
  \BibitemOpen
  \bibfield  {author} {\bibinfo {author} {\bibfnamefont {J.}~\bibnamefont
  {Hwang}}, \bibinfo {author} {\bibfnamefont {M.}~\bibnamefont {Pototschnig}},
  \bibinfo {author} {\bibfnamefont {R.}~\bibnamefont {Lettow}}, \bibinfo
  {author} {\bibfnamefont {G.}~\bibnamefont {Zumofen}}, \bibinfo {author}
  {\bibfnamefont {A.}~\bibnamefont {Renn}}, \bibinfo {author} {\bibfnamefont
  {S.}~\bibnamefont {G{\"{o}}tzinger}},\ and\ \bibinfo {author} {\bibfnamefont
  {V.}~\bibnamefont {Sandoghdar}},\ }\bibfield  {title} {\bibinfo {title} {{A
  single-molecule optical transistor}},\ }\href
  {https://doi.org/10.1038/nature08134} {\bibfield  {journal} {\bibinfo
  {journal} {Nature}\ }\textbf {\bibinfo {volume} {460}},\ \bibinfo {pages}
  {76} (\bibinfo {year} {2009})}\BibitemShut {NoStop}%
\bibitem [{\citenamefont {Tiecke}\ \emph {et~al.}(2014)\citenamefont {Tiecke},
  \citenamefont {Thompson}, \citenamefont {{De Leon}}, \citenamefont {Liu},
  \citenamefont {Vuleti{\'{c}}},\ and\ \citenamefont {Lukin}}]{Tiecke2014}%
  \BibitemOpen
  \bibfield  {author} {\bibinfo {author} {\bibfnamefont {T.~G.}\ \bibnamefont
  {Tiecke}}, \bibinfo {author} {\bibfnamefont {J.~D.}\ \bibnamefont
  {Thompson}}, \bibinfo {author} {\bibfnamefont {N.~P.}\ \bibnamefont {{De
  Leon}}}, \bibinfo {author} {\bibfnamefont {L.~R.}\ \bibnamefont {Liu}},
  \bibinfo {author} {\bibfnamefont {V.}~\bibnamefont {Vuleti{\'{c}}}},\ and\
  \bibinfo {author} {\bibfnamefont {M.~D.}\ \bibnamefont {Lukin}},\ }\bibfield
  {title} {\bibinfo {title} {{Nanophotonic quantum phase switch with a single
  atom}},\ }\href {https://doi.org/10.1038/nature13188} {\bibfield  {journal}
  {\bibinfo  {journal} {Nature}\ }\textbf {\bibinfo {volume} {508}},\ \bibinfo
  {pages} {241} (\bibinfo {year} {2014})},\ \Eprint
  {https://arxiv.org/abs/1404.5615} {arXiv:1404.5615} \BibitemShut {NoStop}%
\bibitem [{\citenamefont {Bhaskar}\ \emph {et~al.}(2020)\citenamefont
  {Bhaskar}, \citenamefont {Riedinger}, \citenamefont {Machielse},
  \citenamefont {Levonian}, \citenamefont {Nguyen}, \citenamefont {Knall},
  \citenamefont {Park}, \citenamefont {Englund}, \citenamefont {Lon{\v{c}}ar},
  \citenamefont {Sukachev},\ and\ \citenamefont {Lukin}}]{Bhaskar2020}%
  \BibitemOpen
  \bibfield  {author} {\bibinfo {author} {\bibfnamefont {M.~K.}\ \bibnamefont
  {Bhaskar}}, \bibinfo {author} {\bibfnamefont {R.}~\bibnamefont {Riedinger}},
  \bibinfo {author} {\bibfnamefont {B.}~\bibnamefont {Machielse}}, \bibinfo
  {author} {\bibfnamefont {D.~S.}\ \bibnamefont {Levonian}}, \bibinfo {author}
  {\bibfnamefont {C.~T.}\ \bibnamefont {Nguyen}}, \bibinfo {author}
  {\bibfnamefont {E.~N.}\ \bibnamefont {Knall}}, \bibinfo {author}
  {\bibfnamefont {H.}~\bibnamefont {Park}}, \bibinfo {author} {\bibfnamefont
  {D.}~\bibnamefont {Englund}}, \bibinfo {author} {\bibfnamefont
  {M.}~\bibnamefont {Lon{\v{c}}ar}}, \bibinfo {author} {\bibfnamefont {D.~D.}\
  \bibnamefont {Sukachev}},\ and\ \bibinfo {author} {\bibfnamefont {M.~D.}\
  \bibnamefont {Lukin}},\ }\bibfield  {title} {\bibinfo {title} {{Experimental
  demonstration of memory-enhanced quantum communication}},\ }\href
  {https://doi.org/10.1038/s41586-020-2103-5} {\bibfield  {journal} {\bibinfo
  {journal} {Nature}\ }\textbf {\bibinfo {volume} {580}},\ \bibinfo {pages}
  {60} (\bibinfo {year} {2020})},\ \Eprint {https://arxiv.org/abs/1909.01323}
  {arXiv:1909.01323} \BibitemShut {NoStop}%
\bibitem [{\citenamefont {Zumofen}\ \emph {et~al.}(2008)\citenamefont
  {Zumofen}, \citenamefont {Mojarad}, \citenamefont {Sandoghdar},\ and\
  \citenamefont {Agio}}]{Zumofen2008}%
  \BibitemOpen
  \bibfield  {author} {\bibinfo {author} {\bibfnamefont {G.}~\bibnamefont
  {Zumofen}}, \bibinfo {author} {\bibfnamefont {N.~M.}\ \bibnamefont
  {Mojarad}}, \bibinfo {author} {\bibfnamefont {V.}~\bibnamefont
  {Sandoghdar}},\ and\ \bibinfo {author} {\bibfnamefont {M.}~\bibnamefont
  {Agio}},\ }\bibfield  {title} {\bibinfo {title} {{Perfect reflection of light
  by an oscillating dipole}},\ }\href
  {https://doi.org/10.1103/PhysRevLett.101.180404} {\bibfield  {journal}
  {\bibinfo  {journal} {Physical Review Letters}\ }\textbf {\bibinfo {volume}
  {101}},\ \bibinfo {pages} {180404} (\bibinfo {year} {2008})},\ \Eprint
  {https://arxiv.org/abs/0805.3231} {arXiv:0805.3231} \BibitemShut {NoStop}%
\bibitem [{\citenamefont {Wrigge}\ \emph {et~al.}(2008)\citenamefont {Wrigge},
  \citenamefont {Gerhardt}, \citenamefont {Hwang}, \citenamefont {Zumofen},\
  and\ \citenamefont {Sandoghdar}}]{Wrigge2008}%
  \BibitemOpen
  \bibfield  {author} {\bibinfo {author} {\bibfnamefont {G.}~\bibnamefont
  {Wrigge}}, \bibinfo {author} {\bibfnamefont {I.}~\bibnamefont {Gerhardt}},
  \bibinfo {author} {\bibfnamefont {J.}~\bibnamefont {Hwang}}, \bibinfo
  {author} {\bibfnamefont {G.}~\bibnamefont {Zumofen}},\ and\ \bibinfo {author}
  {\bibfnamefont {V.}~\bibnamefont {Sandoghdar}},\ }\bibfield  {title}
  {\bibinfo {title} {{Efficient coupling of photons to a single molecule and
  the observation of its resonance fluorescence}},\ }\href
  {https://doi.org/10.1038/nphys812} {\bibfield  {journal} {\bibinfo  {journal}
  {Nature Physics}\ }\textbf {\bibinfo {volume} {4}},\ \bibinfo {pages} {60}
  (\bibinfo {year} {2008})},\ \Eprint {https://arxiv.org/abs/0707.3398}
  {arXiv:0707.3398} \BibitemShut {NoStop}%
\bibitem [{\citenamefont {Sipahigil}\ \emph {et~al.}(2016)\citenamefont
  {Sipahigil}, \citenamefont {Evans}, \citenamefont {Sukachev}, \citenamefont
  {Burek}, \citenamefont {Borregaard}, \citenamefont {Bhaskar}, \citenamefont
  {Nguyen}, \citenamefont {Pacheco}, \citenamefont {Atikian}, \citenamefont
  {Meuwly}, \citenamefont {Camacho}, \citenamefont {Jelezko}, \citenamefont
  {Bielejec}, \citenamefont {Park}, \citenamefont {Lon{\v{c}}ar},\ and\
  \citenamefont {Lukin}}]{Sipahigil2016}%
  \BibitemOpen
  \bibfield  {author} {\bibinfo {author} {\bibfnamefont {A.}~\bibnamefont
  {Sipahigil}}, \bibinfo {author} {\bibfnamefont {R.~E.}\ \bibnamefont
  {Evans}}, \bibinfo {author} {\bibfnamefont {D.~D.}\ \bibnamefont {Sukachev}},
  \bibinfo {author} {\bibfnamefont {M.~J.}\ \bibnamefont {Burek}}, \bibinfo
  {author} {\bibfnamefont {J.}~\bibnamefont {Borregaard}}, \bibinfo {author}
  {\bibfnamefont {M.~K.}\ \bibnamefont {Bhaskar}}, \bibinfo {author}
  {\bibfnamefont {C.~T.}\ \bibnamefont {Nguyen}}, \bibinfo {author}
  {\bibfnamefont {J.~L.}\ \bibnamefont {Pacheco}}, \bibinfo {author}
  {\bibfnamefont {H.~A.}\ \bibnamefont {Atikian}}, \bibinfo {author}
  {\bibfnamefont {C.}~\bibnamefont {Meuwly}}, \bibinfo {author} {\bibfnamefont
  {R.~M.}\ \bibnamefont {Camacho}}, \bibinfo {author} {\bibfnamefont
  {F.}~\bibnamefont {Jelezko}}, \bibinfo {author} {\bibfnamefont
  {E.}~\bibnamefont {Bielejec}}, \bibinfo {author} {\bibfnamefont
  {H.}~\bibnamefont {Park}}, \bibinfo {author} {\bibfnamefont {M.}~\bibnamefont
  {Lon{\v{c}}ar}},\ and\ \bibinfo {author} {\bibfnamefont {M.~D.}\ \bibnamefont
  {Lukin}},\ }\bibfield  {title} {\bibinfo {title} {{An integrated diamond
  nanophotonics platform for quantum-optical networks}},\ }\href
  {https://doi.org/10.1126/science.aah6875} {\bibfield  {journal} {\bibinfo
  {journal} {Science}\ }\textbf {\bibinfo {volume} {354}},\ \bibinfo {pages}
  {847} (\bibinfo {year} {2016})},\ \Eprint {https://arxiv.org/abs/1608.05147}
  {arXiv:1608.05147} \BibitemShut {NoStop}%
\bibitem [{\citenamefont {Wang}\ \emph
  {et~al.}(2019{\natexlab{b}})\citenamefont {Wang}, \citenamefont {Kelkar},
  \citenamefont {Martin-Cano}, \citenamefont {Rattenbacher}, \citenamefont
  {Shkarin}, \citenamefont {Utikal}, \citenamefont {G{\"{o}}tzinger},\ and\
  \citenamefont {Sandoghdar}}]{Wang2018a}%
  \BibitemOpen
  \bibfield  {author} {\bibinfo {author} {\bibfnamefont {D.}~\bibnamefont
  {Wang}}, \bibinfo {author} {\bibfnamefont {H.}~\bibnamefont {Kelkar}},
  \bibinfo {author} {\bibfnamefont {D.}~\bibnamefont {Martin-Cano}}, \bibinfo
  {author} {\bibfnamefont {D.}~\bibnamefont {Rattenbacher}}, \bibinfo {author}
  {\bibfnamefont {A.}~\bibnamefont {Shkarin}}, \bibinfo {author} {\bibfnamefont
  {T.}~\bibnamefont {Utikal}}, \bibinfo {author} {\bibfnamefont
  {S.}~\bibnamefont {G{\"{o}}tzinger}},\ and\ \bibinfo {author} {\bibfnamefont
  {V.}~\bibnamefont {Sandoghdar}},\ }\bibfield  {title} {\bibinfo {title}
  {{Turning a molecule into a coherent two-level quantum system}},\ }\href
  {https://doi.org/10.1038/s41567-019-0436-5} {\bibfield  {journal} {\bibinfo
  {journal} {Nature Physics}\ }\textbf {\bibinfo {volume} {15}},\ \bibinfo
  {pages} {483} (\bibinfo {year} {2019}{\natexlab{b}})}\BibitemShut {NoStop}%
\bibitem [{\citenamefont {Moerner}\ and\ \citenamefont
  {Kador}(1989)}]{Moerner1989}%
  \BibitemOpen
  \bibfield  {author} {\bibinfo {author} {\bibfnamefont {W.~E.}\ \bibnamefont
  {Moerner}}\ and\ \bibinfo {author} {\bibfnamefont {L.}~\bibnamefont
  {Kador}},\ }\bibfield  {title} {\bibinfo {title} {{Optical detection and
  spectroscopy of single molecules in a solid}},\ }\href
  {https://doi.org/10.1103/PhysRevLett.62.2535} {\bibfield  {journal} {\bibinfo
   {journal} {Physical Review Letters}\ }\textbf {\bibinfo {volume} {62}},\
  \bibinfo {pages} {2535} (\bibinfo {year} {1989})}\BibitemShut {NoStop}%
\bibitem [{\citenamefont {Brunel}\ \emph {et~al.}(1999)\citenamefont {Brunel},
  \citenamefont {Lounis}, \citenamefont {Tamarat},\ and\ \citenamefont
  {Orrit}}]{Brunel1999a}%
  \BibitemOpen
  \bibfield  {author} {\bibinfo {author} {\bibfnamefont {C.}~\bibnamefont
  {Brunel}}, \bibinfo {author} {\bibfnamefont {B.}~\bibnamefont {Lounis}},
  \bibinfo {author} {\bibfnamefont {P.}~\bibnamefont {Tamarat}},\ and\ \bibinfo
  {author} {\bibfnamefont {M.}~\bibnamefont {Orrit}},\ }\bibfield  {title}
  {\bibinfo {title} {{Triggered source of single photons based on controlled
  single molecule fluorescence}},\ }\href
  {https://doi.org/10.1103/PhysRevLett.83.2722} {\bibfield  {journal} {\bibinfo
   {journal} {Physical Review Letters}\ }\textbf {\bibinfo {volume} {83}},\
  \bibinfo {pages} {2722} (\bibinfo {year} {1999})}\BibitemShut {NoStop}%
\bibitem [{\citenamefont {Rattenbacher}\ \emph {et~al.}(2019)\citenamefont
  {Rattenbacher}, \citenamefont {Shkarin}, \citenamefont {Renger},
  \citenamefont {Utikal}, \citenamefont {G{\"{o}}tzinger},\ and\ \citenamefont
  {Sandoghdar}}]{Rattenbacher2019}%
  \BibitemOpen
  \bibfield  {author} {\bibinfo {author} {\bibfnamefont {D.}~\bibnamefont
  {Rattenbacher}}, \bibinfo {author} {\bibfnamefont {A.}~\bibnamefont
  {Shkarin}}, \bibinfo {author} {\bibfnamefont {J.}~\bibnamefont {Renger}},
  \bibinfo {author} {\bibfnamefont {T.}~\bibnamefont {Utikal}}, \bibinfo
  {author} {\bibfnamefont {S.}~\bibnamefont {G{\"{o}}tzinger}},\ and\ \bibinfo
  {author} {\bibfnamefont {V.}~\bibnamefont {Sandoghdar}},\ }\bibfield  {title}
  {\bibinfo {title} {{Coherent coupling of single molecules to on-chip ring
  resonators}},\ }\href {https://doi.org/10.1088/1367-2630/ab28b2} {\bibfield
  {journal} {\bibinfo  {journal} {New Journal of Physics}\ }\textbf {\bibinfo
  {volume} {21}},\ \bibinfo {pages} {062002} (\bibinfo {year}
  {2019})}\BibitemShut {NoStop}%
\bibitem [{\citenamefont {T{\"{u}}rschmann}\ \emph {et~al.}(2017)\citenamefont
  {T{\"{u}}rschmann}, \citenamefont {Rotenberg}, \citenamefont {Renger},
  \citenamefont {Harder}, \citenamefont {Lohse}, \citenamefont {Utikal},
  \citenamefont {G{\"{o}}tzinger},\ and\ \citenamefont
  {Sandoghdar}}]{Turschmann2017}%
  \BibitemOpen
  \bibfield  {author} {\bibinfo {author} {\bibfnamefont {P.}~\bibnamefont
  {T{\"{u}}rschmann}}, \bibinfo {author} {\bibfnamefont {N.}~\bibnamefont
  {Rotenberg}}, \bibinfo {author} {\bibfnamefont {J.}~\bibnamefont {Renger}},
  \bibinfo {author} {\bibfnamefont {I.}~\bibnamefont {Harder}}, \bibinfo
  {author} {\bibfnamefont {O.}~\bibnamefont {Lohse}}, \bibinfo {author}
  {\bibfnamefont {T.}~\bibnamefont {Utikal}}, \bibinfo {author} {\bibfnamefont
  {S.}~\bibnamefont {G{\"{o}}tzinger}},\ and\ \bibinfo {author} {\bibfnamefont
  {V.}~\bibnamefont {Sandoghdar}},\ }\bibfield  {title} {\bibinfo {title}
  {{On-chip linear and nonlinear control of single molecules coupled to a
  nanoguide}},\ }\href {https://doi.org/10.1021/acs.nanolett.7b02033}
  {\bibfield  {journal} {\bibinfo  {journal} {Nano Letters}\ }\textbf {\bibinfo
  {volume} {17}},\ \bibinfo {pages} {4941} (\bibinfo {year} {2017})},\ \Eprint
  {https://arxiv.org/abs/1702.05923} {arXiv:1702.05923} \BibitemShut {NoStop}%
\bibitem [{\citenamefont {Lombardi}\ \emph {et~al.}(2018)\citenamefont
  {Lombardi}, \citenamefont {Ovvyan}, \citenamefont {Pazzagli}, \citenamefont
  {Mazzamuto}, \citenamefont {Kewes}, \citenamefont {Neitzke}, \citenamefont
  {Gruhler}, \citenamefont {Benson}, \citenamefont {Pernice}, \citenamefont
  {Cataliotti},\ and\ \citenamefont {Toninelli}}]{Lombardi2018}%
  \BibitemOpen
  \bibfield  {author} {\bibinfo {author} {\bibfnamefont {P.}~\bibnamefont
  {Lombardi}}, \bibinfo {author} {\bibfnamefont {A.~P.}\ \bibnamefont
  {Ovvyan}}, \bibinfo {author} {\bibfnamefont {S.}~\bibnamefont {Pazzagli}},
  \bibinfo {author} {\bibfnamefont {G.}~\bibnamefont {Mazzamuto}}, \bibinfo
  {author} {\bibfnamefont {G.}~\bibnamefont {Kewes}}, \bibinfo {author}
  {\bibfnamefont {O.}~\bibnamefont {Neitzke}}, \bibinfo {author} {\bibfnamefont
  {N.}~\bibnamefont {Gruhler}}, \bibinfo {author} {\bibfnamefont
  {O.}~\bibnamefont {Benson}}, \bibinfo {author} {\bibfnamefont {W.~H.}\
  \bibnamefont {Pernice}}, \bibinfo {author} {\bibfnamefont {F.~S.}\
  \bibnamefont {Cataliotti}},\ and\ \bibinfo {author} {\bibfnamefont
  {C.}~\bibnamefont {Toninelli}},\ }\bibfield  {title} {\bibinfo {title}
  {{Photostable Molecules on Chip: Integrated Sources of Nonclassical Light}},\
  }\href {https://doi.org/10.1021/acsphotonics.7b00521} {\bibfield  {journal}
  {\bibinfo  {journal} {ACS Photonics}\ }\textbf {\bibinfo {volume} {5}},\
  \bibinfo {pages} {126} (\bibinfo {year} {2018})}\BibitemShut {NoStop}%
\bibitem [{\citenamefont {Grandi}\ \emph {et~al.}(2019)\citenamefont {Grandi},
  \citenamefont {Nielsen}, \citenamefont {Cambiasso}, \citenamefont {Boissier},
  \citenamefont {Major}, \citenamefont {Reardon}, \citenamefont {Krauss},
  \citenamefont {Oulton}, \citenamefont {Hinds},\ and\ \citenamefont
  {Clark}}]{Grandi2019}%
  \BibitemOpen
  \bibfield  {author} {\bibinfo {author} {\bibfnamefont {S.}~\bibnamefont
  {Grandi}}, \bibinfo {author} {\bibfnamefont {M.~P.}\ \bibnamefont {Nielsen}},
  \bibinfo {author} {\bibfnamefont {J.}~\bibnamefont {Cambiasso}}, \bibinfo
  {author} {\bibfnamefont {S.}~\bibnamefont {Boissier}}, \bibinfo {author}
  {\bibfnamefont {K.~D.}\ \bibnamefont {Major}}, \bibinfo {author}
  {\bibfnamefont {C.}~\bibnamefont {Reardon}}, \bibinfo {author} {\bibfnamefont
  {T.~F.}\ \bibnamefont {Krauss}}, \bibinfo {author} {\bibfnamefont {R.~F.}\
  \bibnamefont {Oulton}}, \bibinfo {author} {\bibfnamefont {E.~A.}\
  \bibnamefont {Hinds}},\ and\ \bibinfo {author} {\bibfnamefont {A.~S.}\
  \bibnamefont {Clark}},\ }\bibfield  {title} {\bibinfo {title} {{Hybrid
  plasmonic waveguide coupling of photons from a single molecule}},\ }\href
  {https://doi.org/10.1063/1.5110275} {\bibfield  {journal} {\bibinfo
  {journal} {APL Photonics}\ }\textbf {\bibinfo {volume} {4}},\ \bibinfo
  {pages} {086101} (\bibinfo {year} {2019})},\ \Eprint
  {https://arxiv.org/abs/1905.06321} {arXiv:1905.06321} \BibitemShut {NoStop}%
\bibitem [{\citenamefont {Dung}\ \emph {et~al.}(2002)\citenamefont {Dung},
  \citenamefont {Kn{\"{o}}ll},\ and\ \citenamefont {Welsch}}]{Dung2002}%
  \BibitemOpen
  \bibfield  {author} {\bibinfo {author} {\bibfnamefont {H.~T.}\ \bibnamefont
  {Dung}}, \bibinfo {author} {\bibfnamefont {L.}~\bibnamefont {Kn{\"{o}}ll}},\
  and\ \bibinfo {author} {\bibfnamefont {D.-G.}\ \bibnamefont {Welsch}},\
  }\bibfield  {title} {\bibinfo {title} {{Resonant dipole-dipole interaction in
  the presence of dispersing and absorbing surroundings}},\ }\href
  {https://doi.org/10.1103/PhysRevA.66.063810} {\bibfield  {journal} {\bibinfo
  {journal} {Physical Review A - Atomic, Molecular, and Optical Physics}\
  }\textbf {\bibinfo {volume} {66}},\ \bibinfo {pages} {16} (\bibinfo {year}
  {2002})},\ \Eprint {https://arxiv.org/abs/0205056} {arXiv:0205056 [quant-ph]}
  \BibitemShut {NoStop}%
\bibitem [{\citenamefont {Allen}\ and\ \citenamefont
  {Eberly}(1987)}]{Allen1987}%
  \BibitemOpen
  \bibfield  {author} {\bibinfo {author} {\bibfnamefont {L.}~\bibnamefont
  {Allen}}\ and\ \bibinfo {author} {\bibfnamefont {J.~H.}\ \bibnamefont
  {Eberly}},\ }\href@noop {} {\emph {\bibinfo {title} {{Optical Resonance and
  Two-level Atoms}}}}\ (\bibinfo  {publisher} {Dover publications},\ \bibinfo
  {year} {1987})\BibitemShut {NoStop}%
\bibitem [{\citenamefont {Asenjo-Garcia}\ \emph {et~al.}(2017)\citenamefont
  {Asenjo-Garcia}, \citenamefont {Hood}, \citenamefont {Chang},\ and\
  \citenamefont {Kimble}}]{Asenjo-Garcia2017}%
  \BibitemOpen
  \bibfield  {author} {\bibinfo {author} {\bibfnamefont {A.}~\bibnamefont
  {Asenjo-Garcia}}, \bibinfo {author} {\bibfnamefont {J.~D.}\ \bibnamefont
  {Hood}}, \bibinfo {author} {\bibfnamefont {D.~E.}\ \bibnamefont {Chang}},\
  and\ \bibinfo {author} {\bibfnamefont {H.~J.}\ \bibnamefont {Kimble}},\
  }\bibfield  {title} {\bibinfo {title} {{Atom-light interactions in
  quasi-one-dimensional nanostructures: A Green's-function perspective}},\
  }\href {https://doi.org/10.1103/PhysRevA.95.033818} {\bibfield  {journal}
  {\bibinfo  {journal} {Physical Review A}\ }\textbf {\bibinfo {volume} {95}},\
  \bibinfo {pages} {033818} (\bibinfo {year} {2017})},\ \Eprint
  {https://arxiv.org/abs/1606.04977} {arXiv:1606.04977} \BibitemShut {NoStop}%
\bibitem [{\citenamefont {Nicolet}\ \emph
  {et~al.}(2007{\natexlab{a}})\citenamefont {Nicolet}, \citenamefont {Hofmann},
  \citenamefont {Kol'chenko}, \citenamefont {Kozankiewicz},\ and\ \citenamefont
  {Orrit}}]{Nicolet2007a}%
  \BibitemOpen
  \bibfield  {author} {\bibinfo {author} {\bibfnamefont {A.~A.}\ \bibnamefont
  {Nicolet}}, \bibinfo {author} {\bibfnamefont {C.}~\bibnamefont {Hofmann}},
  \bibinfo {author} {\bibfnamefont {M.~A.}\ \bibnamefont {Kol'chenko}},
  \bibinfo {author} {\bibfnamefont {B.}~\bibnamefont {Kozankiewicz}},\ and\
  \bibinfo {author} {\bibfnamefont {M.}~\bibnamefont {Orrit}},\ }\bibfield
  {title} {\bibinfo {title} {{Single dibenzoterrylene molecules in an
  anthracene crystal: Spectroscopy and photophysics}},\ }\href
  {https://doi.org/10.1002/cphc.200700091} {\bibfield  {journal} {\bibinfo
  {journal} {ChemPhysChem}\ }\textbf {\bibinfo {volume} {8}},\ \bibinfo {pages}
  {1215} (\bibinfo {year} {2007}{\natexlab{a}})}\BibitemShut {NoStop}%
\bibitem [{\citenamefont {Nicolet}\ \emph {et~al.}(2006)\citenamefont
  {Nicolet}, \citenamefont {Kol'chenko}, \citenamefont {Kozankiewicz},\ and\
  \citenamefont {Orrit}}]{Nicolet2006}%
  \BibitemOpen
  \bibfield  {author} {\bibinfo {author} {\bibfnamefont {A.~A.}\ \bibnamefont
  {Nicolet}}, \bibinfo {author} {\bibfnamefont {M.~A.}\ \bibnamefont
  {Kol'chenko}}, \bibinfo {author} {\bibfnamefont {B.}~\bibnamefont
  {Kozankiewicz}},\ and\ \bibinfo {author} {\bibfnamefont {M.}~\bibnamefont
  {Orrit}},\ }\bibfield  {title} {\bibinfo {title} {{Intermolecular intersystem
  crossing in single-molecule spectroscopy: Terrylene in anthracene crystal}},\
  }\href {https://doi.org/10.1063/1.2184311} {\bibfield  {journal} {\bibinfo
  {journal} {Journal of Chemical Physics}\ }\textbf {\bibinfo {volume} {124}},\
  \bibinfo {pages} {164711} (\bibinfo {year} {2006})}\BibitemShut {NoStop}%
\bibitem [{\citenamefont {Jelezko}\ \emph {et~al.}(1997)\citenamefont
  {Jelezko}, \citenamefont {Lounis},\ and\ \citenamefont
  {Orrit}}]{Jelezko1997}%
  \BibitemOpen
  \bibfield  {author} {\bibinfo {author} {\bibfnamefont {F.}~\bibnamefont
  {Jelezko}}, \bibinfo {author} {\bibfnamefont {B.}~\bibnamefont {Lounis}},\
  and\ \bibinfo {author} {\bibfnamefont {M.}~\bibnamefont {Orrit}},\ }\bibfield
   {title} {\bibinfo {title} {{Pump-probe spectroscopy and photophysical
  properties of single di-benzanthanthrene molecules in a naphthalene
  crystal}},\ }\href {https://doi.org/10.1063/1.474525} {\bibfield  {journal}
  {\bibinfo  {journal} {Journal of Chemical Physics}\ }\textbf {\bibinfo
  {volume} {107}},\ \bibinfo {pages} {1692} (\bibinfo {year}
  {1997})}\BibitemShut {NoStop}%
\bibitem [{\citenamefont {Nicolet}\ \emph
  {et~al.}(2007{\natexlab{b}})\citenamefont {Nicolet}, \citenamefont {Bordat},
  \citenamefont {Hofmann}, \citenamefont {Kol'chenko}, \citenamefont
  {Kozankiewicz}, \citenamefont {Brown},\ and\ \citenamefont
  {Orrit}}]{Nicolet2007b}%
  \BibitemOpen
  \bibfield  {author} {\bibinfo {author} {\bibfnamefont {A.~A.}\ \bibnamefont
  {Nicolet}}, \bibinfo {author} {\bibfnamefont {P.}~\bibnamefont {Bordat}},
  \bibinfo {author} {\bibfnamefont {C.}~\bibnamefont {Hofmann}}, \bibinfo
  {author} {\bibfnamefont {M.~A.}\ \bibnamefont {Kol'chenko}}, \bibinfo
  {author} {\bibfnamefont {B.}~\bibnamefont {Kozankiewicz}}, \bibinfo {author}
  {\bibfnamefont {R.}~\bibnamefont {Brown}},\ and\ \bibinfo {author}
  {\bibfnamefont {M.}~\bibnamefont {Orrit}},\ }\bibfield  {title} {\bibinfo
  {title} {{Single dibenzoterrylene molecules in an anthracene crystal: Main
  insertion sites}},\ }\href {https://doi.org/10.1002/cphc.200700340}
  {\bibfield  {journal} {\bibinfo  {journal} {ChemPhysChem}\ }\textbf {\bibinfo
  {volume} {8}},\ \bibinfo {pages} {1929} (\bibinfo {year}
  {2007}{\natexlab{b}})}\BibitemShut {NoStop}%
\bibitem [{\citenamefont {Trebbia}\ \emph {et~al.}(2009)\citenamefont
  {Trebbia}, \citenamefont {Ruf}, \citenamefont {Tamarat},\ and\ \citenamefont
  {Lounis}}]{Trebbia2009}%
  \BibitemOpen
  \bibfield  {author} {\bibinfo {author} {\bibfnamefont {J.-B.}\ \bibnamefont
  {Trebbia}}, \bibinfo {author} {\bibfnamefont {H.}~\bibnamefont {Ruf}},
  \bibinfo {author} {\bibfnamefont {P.}~\bibnamefont {Tamarat}},\ and\ \bibinfo
  {author} {\bibfnamefont {B.}~\bibnamefont {Lounis}},\ }\bibfield  {title}
  {\bibinfo {title} {{Efficient generation of near infra-red single photons
  from the zero-phonon line of a single molecule}},\ }\href
  {https://doi.org/10.1364/OE.17.023986} {\bibfield  {journal} {\bibinfo
  {journal} {Optics Express}\ }\textbf {\bibinfo {volume} {17}},\ \bibinfo
  {pages} {23986} (\bibinfo {year} {2009})},\ \Eprint
  {https://arxiv.org/abs/1011.6153} {arXiv:1011.6153} \BibitemShut {NoStop}%
\bibitem [{\citenamefont {Maser}\ \emph {et~al.}(2016)\citenamefont {Maser},
  \citenamefont {Gmeiner}, \citenamefont {Utikal}, \citenamefont
  {G{\"{o}}tzinger},\ and\ \citenamefont {Sandoghdar}}]{Maser2016}%
  \BibitemOpen
  \bibfield  {author} {\bibinfo {author} {\bibfnamefont {A.}~\bibnamefont
  {Maser}}, \bibinfo {author} {\bibfnamefont {B.}~\bibnamefont {Gmeiner}},
  \bibinfo {author} {\bibfnamefont {T.}~\bibnamefont {Utikal}}, \bibinfo
  {author} {\bibfnamefont {S.}~\bibnamefont {G{\"{o}}tzinger}},\ and\ \bibinfo
  {author} {\bibfnamefont {V.}~\bibnamefont {Sandoghdar}},\ }\bibfield  {title}
  {\bibinfo {title} {{Few-photon coherent nonlinear optics with a single
  molecule}},\ }\href {https://doi.org/10.1038/nphoton.2016.63} {\bibfield
  {journal} {\bibinfo  {journal} {Nature Photonics}\ }\textbf {\bibinfo
  {volume} {10}},\ \bibinfo {pages} {450} (\bibinfo {year} {2016})},\ \Eprint
  {https://arxiv.org/abs/1509.05216} {arXiv:1509.05216} \BibitemShut {NoStop}%
\bibitem [{\citenamefont {Faez}\ \emph {et~al.}(2014)\citenamefont {Faez},
  \citenamefont {T{\"{u}}rschmann}, \citenamefont {Haakh}, \citenamefont
  {G{\"{o}}tzinger},\ and\ \citenamefont {Sandoghdar}}]{Faez2014}%
  \BibitemOpen
  \bibfield  {author} {\bibinfo {author} {\bibfnamefont {S.}~\bibnamefont
  {Faez}}, \bibinfo {author} {\bibfnamefont {P.}~\bibnamefont
  {T{\"{u}}rschmann}}, \bibinfo {author} {\bibfnamefont {H.~R.}\ \bibnamefont
  {Haakh}}, \bibinfo {author} {\bibfnamefont {S.}~\bibnamefont
  {G{\"{o}}tzinger}},\ and\ \bibinfo {author} {\bibfnamefont {V.}~\bibnamefont
  {Sandoghdar}},\ }\bibfield  {title} {\bibinfo {title} {{Coherent interaction
  of light and single molecules in a dielectric nanoguide}},\ }\href
  {https://doi.org/10.1103/PhysRevLett.113.213601} {\bibfield  {journal}
  {\bibinfo  {journal} {Physical Review Letters}\ }\textbf {\bibinfo {volume}
  {113}},\ \bibinfo {pages} {213601} (\bibinfo {year} {2014})},\ \Eprint
  {https://arxiv.org/abs/1407.2846} {arXiv:1407.2846} \BibitemShut {NoStop}%
\bibitem [{\citenamefont {Gmeiner}\ \emph {et~al.}(2016)\citenamefont
  {Gmeiner}, \citenamefont {Maser}, \citenamefont {Utikal}, \citenamefont
  {G{\"{o}}tzinger},\ and\ \citenamefont {Sandoghdar}}]{Gmeiner2016a}%
  \BibitemOpen
  \bibfield  {author} {\bibinfo {author} {\bibfnamefont {B.}~\bibnamefont
  {Gmeiner}}, \bibinfo {author} {\bibfnamefont {A.}~\bibnamefont {Maser}},
  \bibinfo {author} {\bibfnamefont {T.}~\bibnamefont {Utikal}}, \bibinfo
  {author} {\bibfnamefont {S.}~\bibnamefont {G{\"{o}}tzinger}},\ and\ \bibinfo
  {author} {\bibfnamefont {V.}~\bibnamefont {Sandoghdar}},\ }\bibfield  {title}
  {\bibinfo {title} {{Spectroscopy and microscopy of single molecules in
  nanoscopic channels: spectral behavior vs. confinement depth}},\ }\href
  {https://doi.org/10.1039/c6cp01698g} {\bibfield  {journal} {\bibinfo
  {journal} {Physical Chemistry Chemical Physics}\ }\textbf {\bibinfo {volume}
  {18}},\ \bibinfo {pages} {19588} (\bibinfo {year} {2016})}\BibitemShut
  {NoStop}%
\bibitem [{\citenamefont {Clear}\ \emph {et~al.}(2020)\citenamefont {Clear},
  \citenamefont {Schofield}, \citenamefont {Major}, \citenamefont {Iles-Smith},
  \citenamefont {Clark},\ and\ \citenamefont {McCutcheon}}]{Clear2020}%
  \BibitemOpen
  \bibfield  {author} {\bibinfo {author} {\bibfnamefont {C.}~\bibnamefont
  {Clear}}, \bibinfo {author} {\bibfnamefont {R.~C.}\ \bibnamefont
  {Schofield}}, \bibinfo {author} {\bibfnamefont {K.~D.}\ \bibnamefont
  {Major}}, \bibinfo {author} {\bibfnamefont {J.}~\bibnamefont {Iles-Smith}},
  \bibinfo {author} {\bibfnamefont {A.~S.}\ \bibnamefont {Clark}},\ and\
  \bibinfo {author} {\bibfnamefont {D.~P.}\ \bibnamefont {McCutcheon}},\
  }\bibfield  {title} {\bibinfo {title} {{Phonon-Induced Optical Dephasing in
  Single Organic Molecules}},\ }\href
  {https://doi.org/10.1103/PhysRevLett.124.153602} {\bibfield  {journal}
  {\bibinfo  {journal} {Physical Review Letters}\ }\textbf {\bibinfo {volume}
  {124}},\ \bibinfo {pages} {153602} (\bibinfo {year} {2020})},\ \Eprint
  {https://arxiv.org/abs/2001.04365} {arXiv:2001.04365} \BibitemShut {NoStop}%
\bibitem [{\citenamefont {Grandi}\ \emph {et~al.}(2016)\citenamefont {Grandi},
  \citenamefont {Major}, \citenamefont {Polisseni}, \citenamefont {Boissier},
  \citenamefont {Clark},\ and\ \citenamefont {Hinds}}]{Grandi2016a}%
  \BibitemOpen
  \bibfield  {author} {\bibinfo {author} {\bibfnamefont {S.}~\bibnamefont
  {Grandi}}, \bibinfo {author} {\bibfnamefont {K.~D.}\ \bibnamefont {Major}},
  \bibinfo {author} {\bibfnamefont {C.}~\bibnamefont {Polisseni}}, \bibinfo
  {author} {\bibfnamefont {S.}~\bibnamefont {Boissier}}, \bibinfo {author}
  {\bibfnamefont {A.~S.}\ \bibnamefont {Clark}},\ and\ \bibinfo {author}
  {\bibfnamefont {E.~A.}\ \bibnamefont {Hinds}},\ }\bibfield  {title} {\bibinfo
  {title} {{Quantum dynamics of a driven two-level molecule with variable
  dephasing}},\ }\href {https://doi.org/10.1103/PhysRevA.94.063839} {\bibfield
  {journal} {\bibinfo  {journal} {Physical Review A - Atomic, Molecular, and
  Optical Physics}\ }\textbf {\bibinfo {volume} {94}},\ \bibinfo {pages}
  {063839} (\bibinfo {year} {2016})}\BibitemShut {NoStop}%
\bibitem [{\citenamefont {{Di Falco}}\ \emph {et~al.}(2008)\citenamefont {{Di
  Falco}}, \citenamefont {O'Faolain},\ and\ \citenamefont
  {Krauss}}]{DiFalco2008a}%
  \BibitemOpen
  \bibfield  {author} {\bibinfo {author} {\bibfnamefont {A.}~\bibnamefont {{Di
  Falco}}}, \bibinfo {author} {\bibfnamefont {L.}~\bibnamefont {O'Faolain}},\
  and\ \bibinfo {author} {\bibfnamefont {T.~F.}\ \bibnamefont {Krauss}},\
  }\bibfield  {title} {\bibinfo {title} {{Dispersion control and slow light in
  slotted photonic crystal waveguides}},\ }\href
  {https://doi.org/10.1063/1.2885072} {\bibfield  {journal} {\bibinfo
  {journal} {Applied Physics Letters}\ }\textbf {\bibinfo {volume} {92}},\
  \bibinfo {pages} {083501} (\bibinfo {year} {2008})}\BibitemShut {NoStop}%
\bibitem [{\citenamefont {Muschik}\ \emph {et~al.}(2014)\citenamefont
  {Muschik}, \citenamefont {Moulieras}, \citenamefont {Bachtold}, \citenamefont
  {Koppens}, \citenamefont {Lewenstein},\ and\ \citenamefont
  {Chang}}]{Muschik2014}%
  \BibitemOpen
  \bibfield  {author} {\bibinfo {author} {\bibfnamefont {C.~A.}\ \bibnamefont
  {Muschik}}, \bibinfo {author} {\bibfnamefont {S.}~\bibnamefont {Moulieras}},
  \bibinfo {author} {\bibfnamefont {A.}~\bibnamefont {Bachtold}}, \bibinfo
  {author} {\bibfnamefont {F.~H.}\ \bibnamefont {Koppens}}, \bibinfo {author}
  {\bibfnamefont {M.}~\bibnamefont {Lewenstein}},\ and\ \bibinfo {author}
  {\bibfnamefont {D.~E.}\ \bibnamefont {Chang}},\ }\bibfield  {title} {\bibinfo
  {title} {{Harnessing vacuum forces for quantum sensing of graphene motion}},\
  }\href {https://doi.org/10.1103/PhysRevLett.112.223601} {\bibfield  {journal}
  {\bibinfo  {journal} {Physical Review Letters}\ }\textbf {\bibinfo {volume}
  {112}},\ \bibinfo {pages} {223601} (\bibinfo {year} {2014})},\ \Eprint
  {https://arxiv.org/abs/1304.8090} {arXiv:1304.8090} \BibitemShut {NoStop}%
\bibitem [{\citenamefont {Hettich}\ \emph {et~al.}(2002)\citenamefont
  {Hettich}, \citenamefont {Schmitt}, \citenamefont {Zitzmann}, \citenamefont
  {K{\"{u}}hn}, \citenamefont {Gerhardt},\ and\ \citenamefont
  {Sandoghdar}}]{Hettich2002}%
  \BibitemOpen
  \bibfield  {author} {\bibinfo {author} {\bibfnamefont {C.}~\bibnamefont
  {Hettich}}, \bibinfo {author} {\bibfnamefont {C.}~\bibnamefont {Schmitt}},
  \bibinfo {author} {\bibfnamefont {J.}~\bibnamefont {Zitzmann}}, \bibinfo
  {author} {\bibfnamefont {S.}~\bibnamefont {K{\"{u}}hn}}, \bibinfo {author}
  {\bibfnamefont {I.}~\bibnamefont {Gerhardt}},\ and\ \bibinfo {author}
  {\bibfnamefont {V.}~\bibnamefont {Sandoghdar}},\ }\bibfield  {title}
  {\bibinfo {title} {{Nanometer resolution and coherent optical dipole coupling
  of two individual molecules}},\ }\href
  {https://doi.org/10.1126/science.1075606} {\bibfield  {journal} {\bibinfo
  {journal} {Science}\ }\textbf {\bibinfo {volume} {298}},\ \bibinfo {pages}
  {385} (\bibinfo {year} {2002})}\BibitemShut {NoStop}%
\bibitem [{\citenamefont {Oskooi}\ and\ \citenamefont
  {Johnson}(2013)}]{Oskooi2013}%
  \BibitemOpen
  \bibfield  {author} {\bibinfo {author} {\bibfnamefont {A.}~\bibnamefont
  {Oskooi}}\ and\ \bibinfo {author} {\bibfnamefont {S.~G.}\ \bibnamefont
  {Johnson}},\ }\bibfield  {title} {\bibinfo {title} {{Electromagnetic Wave
  Source Conditions}},\ }\href {http://arxiv.org/abs/1301.5366} {\bibfield
  {journal} {\bibinfo  {journal} {Advances in FDTD Computational
  Electrodynamics: Photonics and Nanotechnology}\ ,\ \bibinfo {pages} {65}}
  (\bibinfo {year} {2013})},\ \Eprint {https://arxiv.org/abs/1301.5366}
  {arXiv:1301.5366} \BibitemShut {NoStop}%
\bibitem [{\citenamefont {Novotny}\ and\ \citenamefont
  {Hecht}(2009)}]{Novotny2009}%
  \BibitemOpen
  \bibfield  {author} {\bibinfo {author} {\bibfnamefont {L.}~\bibnamefont
  {Novotny}}\ and\ \bibinfo {author} {\bibfnamefont {B.}~\bibnamefont
  {Hecht}},\ }\href@noop {} {\emph {\bibinfo {title} {Principles of
  Nano-Optics}}}\ (\bibinfo  {publisher} {Cambridge University Press},\
  \bibinfo {year} {2009})\BibitemShut {NoStop}%
\bibitem [{\citenamefont {Oskooi}\ \emph {et~al.}(2010)\citenamefont {Oskooi},
  \citenamefont {Roundy}, \citenamefont {Ibanescu}, \citenamefont {Bermel},
  \citenamefont {Joannopoulos},\ and\ \citenamefont {Johnson}}]{Oskooi2010}%
  \BibitemOpen
  \bibfield  {author} {\bibinfo {author} {\bibfnamefont {A.~F.}\ \bibnamefont
  {Oskooi}}, \bibinfo {author} {\bibfnamefont {D.}~\bibnamefont {Roundy}},
  \bibinfo {author} {\bibfnamefont {M.}~\bibnamefont {Ibanescu}}, \bibinfo
  {author} {\bibfnamefont {P.}~\bibnamefont {Bermel}}, \bibinfo {author}
  {\bibfnamefont {J.~D.}\ \bibnamefont {Joannopoulos}},\ and\ \bibinfo {author}
  {\bibfnamefont {S.~G.}\ \bibnamefont {Johnson}},\ }\bibfield  {title}
  {\bibinfo {title} {{Meep: A flexible free-software package for
  electromagnetic simulations by the FDTD method}},\ }\href
  {https://doi.org/10.1016/j.cpc.2009.11.008} {\bibfield  {journal} {\bibinfo
  {journal} {Computer Physics Communications}\ }\textbf {\bibinfo {volume}
  {181}},\ \bibinfo {pages} {687} (\bibinfo {year} {2010})}\BibitemShut
  {NoStop}%
\bibitem [{\citenamefont {Yao}\ \emph {et~al.}(2010)\citenamefont {Yao},
  \citenamefont {Mangarao},\ and\ \citenamefont {Hughes}}]{Yao2010}%
  \BibitemOpen
  \bibfield  {author} {\bibinfo {author} {\bibfnamefont {P.}~\bibnamefont
  {Yao}}, \bibinfo {author} {\bibfnamefont {V.~S.}\ \bibnamefont {Mangarao}},\
  and\ \bibinfo {author} {\bibfnamefont {S.}~\bibnamefont {Hughes}},\
  }\bibfield  {title} {\bibinfo {title} {{On-chip single photon sources using
  planar photonic crystals and single quantum dots}},\ }\href
  {https://doi.org/10.1002/lpor.200810081} {\bibfield  {journal} {\bibinfo
  {journal} {Laser and Photonics Reviews}\ }\textbf {\bibinfo {volume} {4}},\
  \bibinfo {pages} {499} (\bibinfo {year} {2010})}\BibitemShut {NoStop}%
\bibitem [{\citenamefont {Chang}\ \emph {et~al.}(2007)\citenamefont {Chang},
  \citenamefont {S{\o}rensen}, \citenamefont {Demler},\ and\ \citenamefont
  {Lukin}}]{Chang2007}%
  \BibitemOpen
  \bibfield  {author} {\bibinfo {author} {\bibfnamefont {D.~E.}\ \bibnamefont
  {Chang}}, \bibinfo {author} {\bibfnamefont {A.~S.}\ \bibnamefont
  {S{\o}rensen}}, \bibinfo {author} {\bibfnamefont {E.~A.}\ \bibnamefont
  {Demler}},\ and\ \bibinfo {author} {\bibfnamefont {M.~D.}\ \bibnamefont
  {Lukin}},\ }\bibfield  {title} {\bibinfo {title} {{A single-photon transistor
  using nanoscale surface plasmons}},\ }\href
  {https://doi.org/10.1038/nphys708} {\bibfield  {journal} {\bibinfo  {journal}
  {Nature Physics}\ }\textbf {\bibinfo {volume} {3}},\ \bibinfo {pages} {807}
  (\bibinfo {year} {2007})},\ \Eprint {https://arxiv.org/abs/0706.4335}
  {arXiv:0706.4335} \BibitemShut {NoStop}%
\bibitem [{\citenamefont {Chang}\ \emph {et~al.}(2012)\citenamefont {Chang},
  \citenamefont {Jiang}, \citenamefont {Gorshkov},\ and\ \citenamefont
  {Kimble}}]{Chang2012}%
  \BibitemOpen
  \bibfield  {author} {\bibinfo {author} {\bibfnamefont {D.~E.}\ \bibnamefont
  {Chang}}, \bibinfo {author} {\bibfnamefont {L.}~\bibnamefont {Jiang}},
  \bibinfo {author} {\bibfnamefont {A.~V.}\ \bibnamefont {Gorshkov}},\ and\
  \bibinfo {author} {\bibfnamefont {H.~J.}\ \bibnamefont {Kimble}},\ }\bibfield
   {title} {\bibinfo {title} {{Cavity QED with atomic mirrors}},\ }\href
  {https://doi.org/10.1088/1367-2630/14/6/063003} {\bibfield  {journal}
  {\bibinfo  {journal} {New Journal of Physics}\ }\textbf {\bibinfo {volume}
  {14}},\ \bibinfo {pages} {063003} (\bibinfo {year} {2012})},\ \Eprint
  {https://arxiv.org/abs/1201.0643} {arXiv:1201.0643} \BibitemShut {NoStop}%
\bibitem [{\citenamefont {Auffeves-Garnier}\ \emph {et~al.}(2006)\citenamefont
  {Auffeves-Garnier}, \citenamefont {Simon}, \citenamefont {Gerard},\ and\
  \citenamefont {Poizat}}]{Auffeves-Garnier2007}%
  \BibitemOpen
  \bibfield  {author} {\bibinfo {author} {\bibfnamefont {A.}~\bibnamefont
  {Auffeves-Garnier}}, \bibinfo {author} {\bibfnamefont {C.}~\bibnamefont
  {Simon}}, \bibinfo {author} {\bibfnamefont {J.-M.}\ \bibnamefont {Gerard}},\
  and\ \bibinfo {author} {\bibfnamefont {J.-P.}\ \bibnamefont {Poizat}},\
  }\bibfield  {title} {\bibinfo {title} {{Giant Optical Non-linearity induced
  by a Single Two-Level System interacting with a Cavity in the Purcell
  Regime}},\ }\href {https://doi.org/10.1103/PhysRevA.75.053823} {\bibfield
  {journal} {\bibinfo  {journal} {Physical Review A - Atomic, Molecular, and
  Optical Physics}\ }\textbf {\bibinfo {volume} {75}},\ \bibinfo {pages}
  {053823} (\bibinfo {year} {2006})},\ \Eprint {https://arxiv.org/abs/0610172}
  {arXiv:0610172 [quant-ph]} \BibitemShut {NoStop}%
\end{thebibliography}
%

~
\clearpage

\onecolumngrid
\begin{center}
  \textbf{\large Supplementary Information for: \protect\\[.1cm] Coherent characterisation of a single molecule in a photonic black box}\\[.2cm]
  Sebastien Boissier,$^1$ Ross C. Schofield,$^1$ Lin Jin,$^2$ Anna Ovvyan,$^2$ Salahuddin Nur,$^1$ Frank H. L. Koppens,$^3$ \\ Costanza Toninelli,$^4$  Wolfram H. P. Pernice,$^2$ Kyle D. Major,$^1$ E. A. Hinds,$^1$ and Alex S. Clark$^{1,*}$\\[.1cm]
  {\itshape ${}^1$Centre for Cold Matter, Blackett Laboratory, Imperial College London,\\ Prince Consort Road, SW7 2AZ, London, United Kingdom\\
  ${}^2$Physikalisches Institut, Westf\"{a}lische Wilhelms, Universit\"{a}t M\"{u}nster, \\Heisenbergstrasse 11, 48149 M\"{u}nster, Germany\\
  ${}^3$ICFO -- Institut de Ciencies Fotoniques, The Barcelona Institute of Science and Technology, \\08860 Castelldefels (Barcelona), Spain\\
  ${^4}$LENS and CNR-INO, Via Nello Carrara 1, 50019 Sesto Fiorentino (FI), Italy}\\
  ${}^*$Email: alex.clark@imperial.ac.uk\\
(Dated: \today)\\[1cm]
\end{center}

\setcounter{page}{1}
\renewcommand\theequation{S\arabic{equation}}
\setcounter{equation}{0}
\renewcommand\thefigure{S\arabic{figure}}
\setcounter{figure}{0}
\renewcommand{\theHtable}{Supplement.\thetable}
\renewcommand{\theHfigure}{Supplement.\thefigure}

\section{Analytical solutions for simple geometries}

For structures where a normal mode decomposition is appropriate, we can compare Eq. 9 and Eq. 10 of the main text with analytical solutions for the Green function, or with results of coupled-mode theory, to obtain expressions that depend explicitly on structural parameters. We do this for the case of a single-mode continuous waveguide and for a cavity in the weak-coupling regime.

\subsection{Continuous waveguide}

For a continuous waveguide with no loss, $|r_{0}|=0$ and $|t_{0}|=1$. In that case, the analytical solution for the Green function \cite{Yao2010} allows us to write Eq. 6 of the main text as \cite{Asenjo-Garcia2017, Turschmann2019}

\begin{equation} \label{wgfield}
    \mathcal{E}_{\text{out}} = \left(1 + i \frac{\beta_{\text{g}} \gamma_{1}}{\Omega} \sigma^{-}  \right) |\mathcal{E}_{\text{in}}| \,,
\end{equation}

\noindent where $\beta_{\text{g}}=2\beta_\text{pump} = 2\beta_\text{probe}$. This is the same result as we would have obtained by applying bosonic annihilation and creation operators to the waveguide modes \cite{Chang2007, Chang2012}. Comparison of  \eqRef{\ref{wgfield}} with Eq. 6 of the main text shows that $\phi_{\text{T}} = \pi/2$ and hence that the normalised transmission given by Eq. 9 of the main text is

\begin{equation}
    \frac{P_{\text{out}}}{P_{\text{in}}} = 1 - \alpha\beta_{\text{g}} \left( 2 -\alpha \beta_{\text{g}} \right) \frac{\Gamma_{1}/(2 \Gamma_{2})}{(\delta\omega / \Gamma_{2})^{2} + 1 + S} \,.
\end{equation}

\noindent The same expression can be found in \cite{Turschmann2019}. Because $|r_{0}|=0$, the reflection spectrum given by Eq. 10 of the main text becomes

\begin{equation}
    \frac{P_{\text{refl}}}{P_{\text{in}}} = (\alpha \beta_{\text{g}})^{2} \frac{\Gamma_{1}/(2 \Gamma_{2})}{(\delta\omega / \Gamma_{2})^{2} + 1 + S} \,.
\end{equation}

\subsection{Symmetrical cavity in the weak-coupling regime}

Let us now consider a waveguide interrupted by a symmetrical cavity, whose mode is matched to the transverse guide mode. In the absence of an emitter the transmission coefficient for the field is (see, e.g. \cite{Auffeves-Garnier2007})

\begin{equation}
    t_{0}=\frac{-1}{1-i(\omega-\omega_C )/\kappa} \,,\\ \label{eq:cavity transmission}
\end{equation}

\noindent where $\omega_{c}$ is the resonant frequency of the cavity and $\kappa$ is the damping rate for the cavity field decaying freely into the two waveguides. When the emitter is pumped by light that is resonant with the cavity, we call the (partial) radiation rate of the excited emitter into the cavity mode $\gamma_{\text{cav}}$. In general, the cavity is not resonant with the pump light and then the rate decreases to $|t_0|^2\gamma_{\text{cav}}$ \cite{Lodahl2015}. Hence, in the language of the main text

\begin{equation}
    \beta_{\text{cav}}\gamma_1=|t_0|^2\gamma_{\text{cav}} \,,\\ \label{eq:gammacav}
\end{equation}

\noindent where $\beta_{\text{cav}}=2\beta_\text{pump} = 2\beta_\text{probe}$.  Neglecting the quantum noise of the field, as in the main text, the literature \cite{Auffeves-Garnier2007, Javadi2015} gives the following relation between the input and output fields in the weak coupling regime:

\begin{equation}
   \mathcal{E}_{\text{out}} = t_{0}\mathcal{E}_{\text{in}}  + i\sqrt{\frac{\gamma_{\text{cav}}}{2}} t_{0} \sigma^{-} \,.\\\label{cav inout}
\end{equation} \

\noindent Reference \cite{Javadi2015} also gives the steady-state solution for the Rabi frequency as

 \begin{equation}
    \Omega = - 2 t_{0}\mathcal{E}_{\text{in}} \sqrt{\frac{\gamma_{\text{cav}}}{2}} \,.\\\label{cavrabi}
\end{equation}

\noindent We substitute  \eqRef{\ref{cavrabi}} and  \eqRef{\ref{eq:gammacav}} into \eqRef{\ref{cav inout}} to find, neglecting an over-all phase, that

\begin{equation}
    \mathcal{E}_{\text{out}} = \left( | t_{0} | - i \frac{t_{0}}{|t_{0}|}\frac{\beta_{\text{cav}}\gamma_{1}}{\Omega} \sigma^- \right) | \mathcal{E}_{\text{in}}  |  \,,\\ \label{caveout}
\end{equation}
\noindent On comparing this result with Eq. 6 of the main text, we see that $\phi_{\text{T}}= \pi/2 + \operatorname{Arg}(-t_{0})$, which depends on the scaled detuning from the cavity resonance $(\omega-\omega_C )/\kappa$, but not on the position of the emitter, as also noted in \cite{Auffeves-Garnier2007}. Following \cite{Auffeves-Garnier2007} this analysis is readily extended to lossy and non-symmetric cavities.

\section{Coupling efficiency simulations}

\begin{figure}[h]
	\begin{center}
	\includegraphics{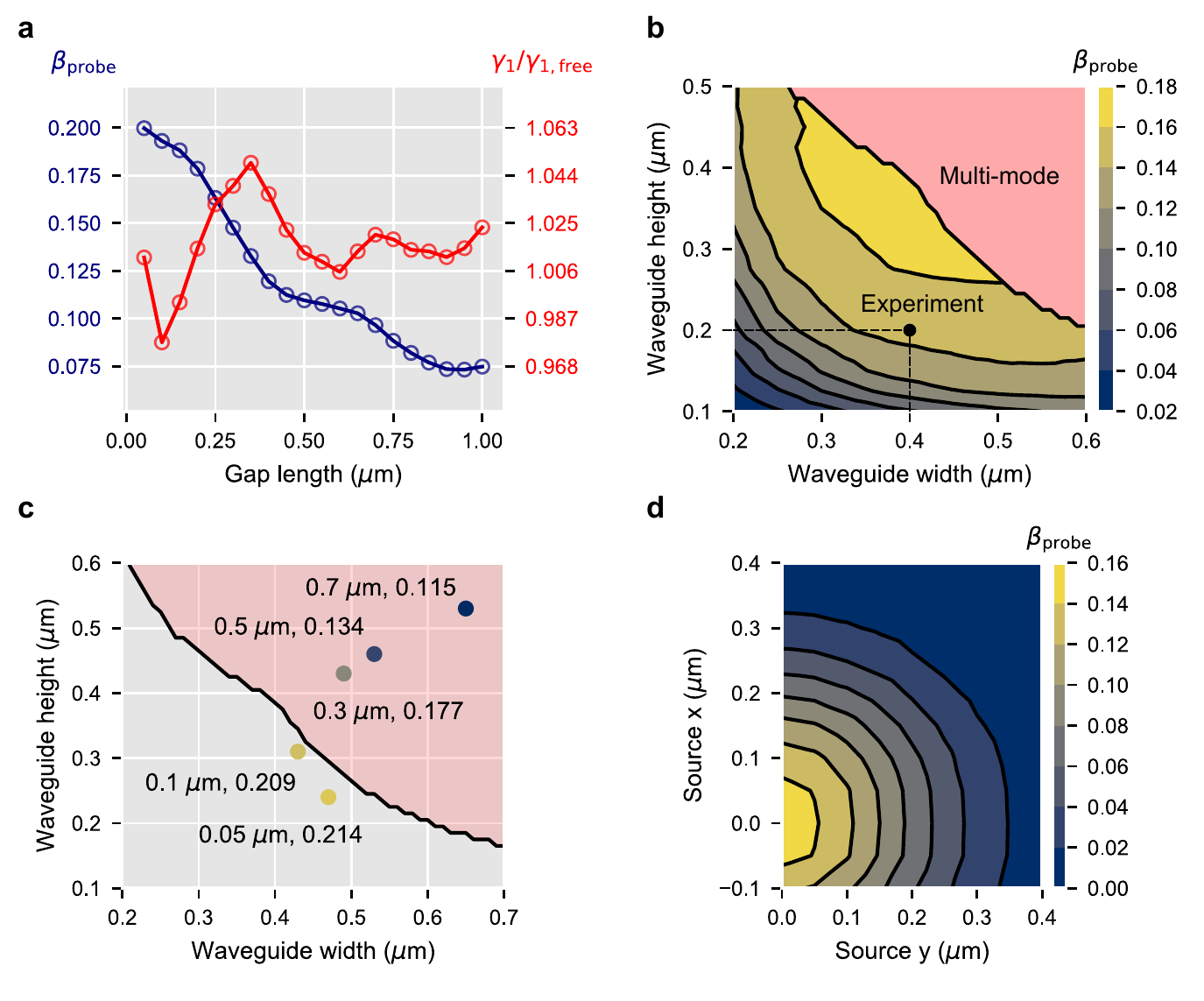}
	\end{center}
	\caption[FDTD simulations to explore the coupling efficiency in our device.]{\textbf{FDTD simulations to explore the coupling efficiency in our device.} \textbf{a}, One-way efficiency $\beta_{\text{probe}}$ and Purcell factor $\gamma_1/\gamma_{\text{1,free}}$ as a function of gap length. The waveguide is \SI{400}{\nano\meter} wide and \SI{200}{\nano\meter} high. The dipole is polarised along the width of the waveguide and lies in the middle of the gap on the centre line of the waveguide.  \textbf{b}, With gap length~=~\SI{300}{\nano\meter}, one-way collection efficiency as a function of waveguide height and width. The waveguide is multi-mode in the pink area. \textbf{c}, The position of each point shows the waveguide dimensions that maximise $\beta_{\text{probe}}$ for a given gap length, and the points are labelled by $(\text{gap length}, \beta_{\text{probe}})$. \textbf{d}, Efficiency $\beta_{\text{probe}}$ as a function of dipole position in the transverse plane at the centre of the \SI{300}{\nano\meter} gap. Waveguide width~=~\SI{400}{\nano\meter}, waveguide height~=~\SI{200}{\nano\meter}.}
	\label{fig:sims}	
\end{figure}

We used finite-difference, time-domain (FDTD) calculations to explore the behaviour of the device and to predict values for the coupling efficiency $\beta_{\text{probe}}$ and phase shift $\phi_{\text{T}}$. The following simulation results are obtained with the channel height and under-etch (defined in Fig. 2(c) of the main text) fixed at \SI{1}{\micro\meter} and \SI{150}{\nano\meter} respectively, which are the values used in the experiment. The channel width is set at the gap length + \SI{200}{\nano\meter} to allow for the uncertainty in alignment. \figRef{\ref{fig:sims}(a-c)} show simulation results for a dipole sitting in the middle of the gap on the centre line of the waveguides, and polarised along $x$. In \figRef{\ref{fig:sims}(a)} we show the coupling efficiency $\beta_{\text{probe}}$ for a waveguide \SI{200}{\nano\meter} high and \SI{400}{\nano\meter} wide, these being the dimensions used in our experiment.  We see that $\beta_{\text{probe}}$ decreases rapidly with the length of the gap and conclude that a short gap is necessary for good coupling. On the other hand, we know that the molecule loses spectral stability if it is less than one or two hundred nm from the interface at the end of the guide, so we consider a good gap length may be \SI{300}{\nano\meter}. The same graph also shows the Purcell factor $\gamma_1/\gamma_{\text{1,free}}$, where $\gamma_{\text{1,free}}$ is the value of $\gamma_1$ when the dipole is in homogeneous anthracene. This factor stays close to 1.

Fixing the gap length at \SI{300}{\nano\meter}, we plot $\beta_{\text{probe}}$ in \figRef{\ref{fig:sims}(b)} as a function of the waveguide height and width. As the size of the guide increases, so does $\beta_{\text{probe}}$; indeed, this plot shows that the coupling efficiency found in \figRef{\ref{fig:sims}(a)} could be improved by increasing the waveguide height to \SI{350}{\nano\meter} or \SI{400}{\nano\meter}. We can understand this behaviour by noting that a larger guided mode diffracts less strongly in the gap and this reduced angular spread increases the overlap of the mode field with the field of the dipole, resulting in an increase in $\beta_{\text{probe}}$. However, if we continue to increase the size of the guide, the increasing spatial spread of the mode starts to reduce the overlap and $\beta_{\text{probe}}$ eventually declines again. In \figRef{\ref{fig:sims}(b)} the guide becomes multi-mode before  the maximum of $\beta_{\text{probe}}$ is reached, and that is the region shown in pink. In this experiment we do not want to operate in the multi-mode regime.

In \figRef{\ref{fig:sims}(c)}, we show for several different gap lengths the waveguide dimensions that maximise $\beta_{\text{probe}}$. With a \SI{300}{\nano\meter} gap the maximum $\beta_{\text{probe}}$ is $17.7\%$ and this lies in the multi-mode regime. On reducing the gap to \SI{100}{\nano\meter} the maximum of $\beta_{\text{probe}}$ moves into the single-mode regime and increases to $20.9\%$, but in practice this gap is too small to expect the molecule to be spectrally stable.

Finally, \figRef{\ref{fig:sims}(d)} shows the variation in $\beta_{\text{probe}}$ when we vary the position of the dipole in the transverse ($xy$) direction so that it no longer sits at the maximum field of the mode in that plane.  The guide dimensions are the same as in \figRef{\ref{fig:sims}(a)} and the gap is \SI{300}{\nano\meter} long.  We see that a transverse displacement away from the axis by up to \SI{100}{\nano\meter} in any direction decreases $\beta_{\text{probe}}$ by less than $4\%$. That is roughly the displacement we deduce for the molecule studied in the main text.

\section{Micro-fluidic channel filling}

 \figRef{\ref{fig:cold_channel}(a)} shows an optical microscope image of unfilled micro-fluidic channels. The channels are \SI{1}{\micro\meter} high and \SI{5}{\micro\meter} wide, tapering to smaller widths in regions where they intersect with the waveguides. Capillary action fills the channels from the edges of the chip with DBT-doped liquid anthracene, which then cools and solidifies. Typically, we find long regions of the solid, separated by shorter voids, as seen in \figRef{\ref{fig:cold_channel}(b)}. We believe this is due to the formation of separate crystals, which shrink away from each other when the anthracene solidifies, making the transition to higher density.

 In \figRef{\ref{fig:cold_channel}(c)} we show the image from a  scanning confocal fluorescence microscope centred on a segment of a channel that tapers down to a width of \SI{2}{\micro\meter}. Operating at \SI{785}{\nano\meter} wavelength, this image reveals three bright fluorescent centres just outside the taper, which correspond to single DBT molecules inside the channel. On changing the wavelength we find more molecules, some inside the \SI{2}{\micro\meter}-wide section, each of which  has a fluorescence spectrum that is typically \SI[parse-numbers=false]{100-200}{\MHz} wide. The blue bars in \figRef{\ref{fig:cold_channel}(d)} show a histogram of the resonance linewidths. We repeated these measurements on molecules in a  \SI{0.5}{\micro\meter}-wide channel and found the distribution of widths plotted in red in \figRef{\ref{fig:cold_channel}(d)}. Because the two distributions are essentially identical, we conclude that the width is not due to spectral instability \cite{Gmeiner2016a} associated with the confining environment, but is simply due to relaxation of the optical dipole through its interaction with thermal phonons. This was to be expected because the temperature of the molecules was \SI{4.7}{\K} - well above the \SI{3}{\K} at which the width normally approaches the limiting value of \SI{40}{\MHz}. In short, the DBT molecules seem well-behaved even inside the small capillaries.

 Finally, we varied the angle of the linearly-polarised pump light at low saturation, and recorded the intensity of the fluorescence as a function of the angle. The circles in \figRef{\ref{fig:cold_channel}(e)} show data for a typical single molecule, with the solid line showing an excellent fit to the expected $\operatorname{cos}^{2}(\theta)$ dependence. Here $\theta$ is the angle between the laser polarisation and the linear transition dipole-moment of the molecule. In bulk crystals, the orientation of the dipole moment is along the b-axis of the anthracene crystal \cite{Nicolet2007b}. We checked to see if the molecules in a \SI{0.5}{\micro\meter} channel are similarly aligned by recording fluorescence spectra of many molecules, each at a range of laser polarisation angles. In \figRef{\ref{fig:cold_channel}(f)} we plot a histogram showing how the orientation of the optical dipoles is distributed around the mean, which we take as \SI{0}{\degree}. We see that the DBT molecules do indeed have a preferred orientation, which we presume is the b-axis of that particular crystal. This result, together with the good spectral stability, confirms that the melt growth produces crystalline anthracene inside the channels.

\section{Imperfect filtering of off-resonant light}

In Eq. 11 of the main text we give an expression for the power transmitted through the waveguides to the detector, normalised to 1 far from resonance. We reproduce that equation here:
\begin{equation}
        T =   1 - \frac{\alpha \beta_{\text{eff}}}{|t_{0}|} \bigg\{ 2 \left( \sin(\phi_{\text{T}}) + \frac{\delta\omega}{\Gamma_{2}} \cos(\phi_{\text{T}})  \right) - \frac{ \alpha \beta_{\text{eff}}}{|t_{0}|} \bigg\} \frac{\Gamma_{1}/(2 \Gamma_{2})}{(\delta\omega / \Gamma_{2})^{2} + 1 + S} \, .
 	\label{eq:main11}
\end{equation}

\noindent The fourth (i.e. the last) term represents the light at frequency $\omega$ that is scattered by the emitter into the probe guide and makes its way through the optics to the detector, which we will call $P(\omega)$. Here we consider the effect of light scattered at other frequencies due to any radiative sidebands that there may be. Let the power reaching the detector in these sidebands be $P(\omega') =\epsilon P(\omega)$. This causes \eqRef{\ref{eq:main11}} to become

\begin{equation}
        T =   1 - \frac{\alpha \beta_{\text{eff}}}{|t_{0}|} \bigg\{ 2 \left( \sin(\phi_{\text{T}}) + \frac{\delta\omega}{\Gamma_{2}} \cos(\phi_{\text{T}})  \right) - \frac{ \alpha \beta_{\text{eff}}}{|t_{0}|} (1+\epsilon)\bigg\} \frac{\Gamma_{1}/(2 \Gamma_{2})}{(\delta\omega / \Gamma_{2})^{2} + 1 + S} \, ,
 	\label{eq:transmission3}
\end{equation}

\noindent where the off-resonant light produces no interference term because it is the time-averaged power that we measure. For most applications it will be desirable to have $\epsilon\ll1$, but even with a good choice of emitter and with filtering of the output, $\epsilon$ may well not be zero. In the case of our experiment with the DBT molecule, we estimate that $\epsilon<5\times10^{-2}$. In order to see the effect of this off-resonant light on our determination of $\beta_{\text{eff}}$ and $\phi_{\text{T}}$, we take a set of values relevant for our experiment: $\alpha=0.33; \beta_{\text{eff}}=0.09; |t_0|=0.7; \phi_{\text{T}}=61\degree; \Gamma_{1}/(2 \Gamma_{2})=0.25$. With these values we make a synthetic data set for $T$ using \eqRef{\ref{eq:transmission3}}, in which we take $\epsilon=1$. When we fit \eqRef{\ref{eq:main11}} to the synthetic data, taking $\beta_{\text{eff}}$ and $\phi_{\text{T}}$ as the fit parameters, the effect of the greatly exaggerated off-resonant light is to change $\beta$ from $9\%$ to $8.8\%$ and to change $\phi_{\text{T}}$ from $61\degree$ to $60.4\degree$. We conclude that the small amount of off-resonant light that may be reaching the detector in our experiment has a negligible effect on the determination of $\beta_{\text{eff}}$ and $\phi_{\text{T}}$.

\newpage
~

\vspace*{\fill}

\vspace*{\fill}

\newpage
~
\vspace*{\fill}

\begin{figure*}[h]
	\begin{center}
	\includegraphics{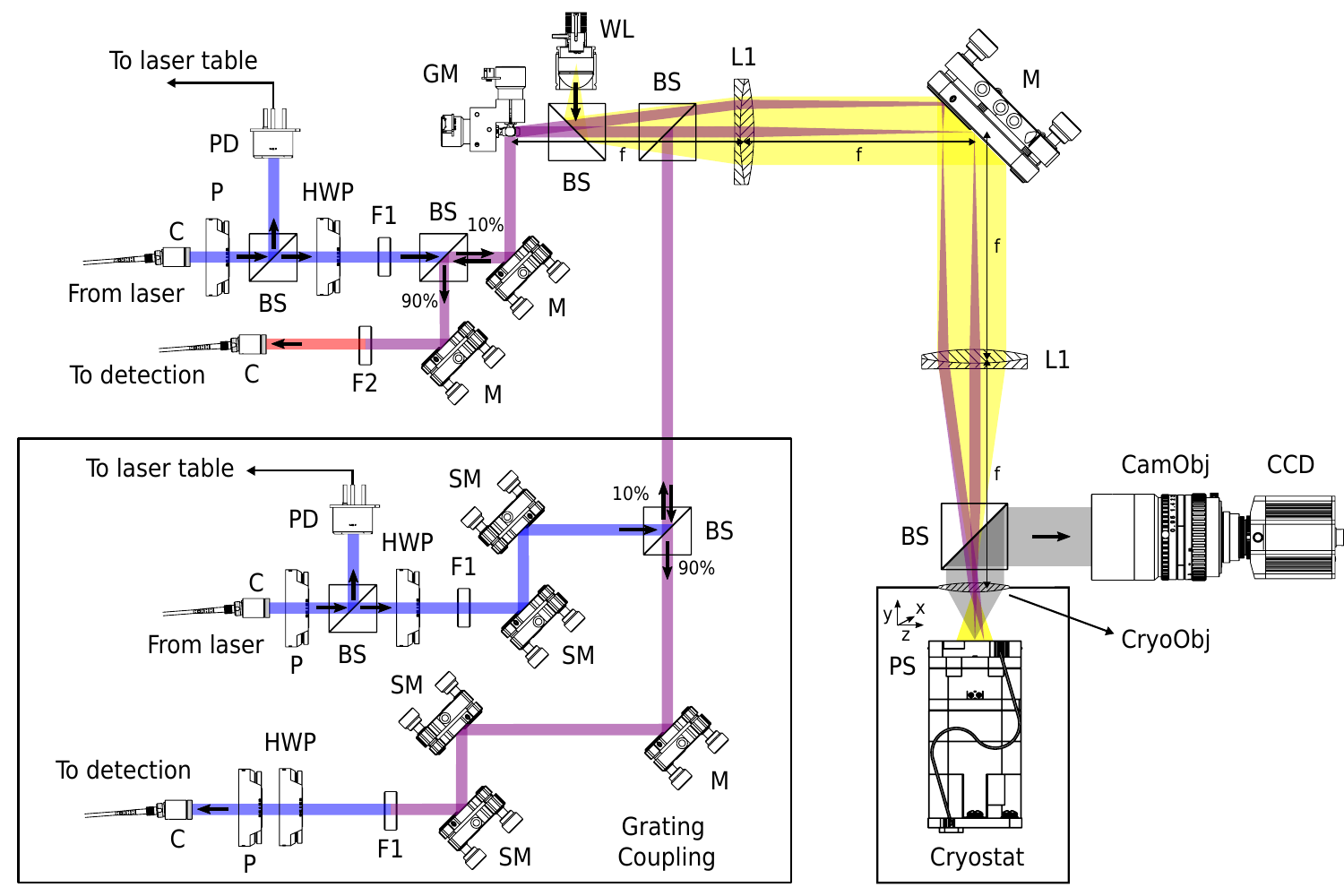}
	\end{center}
	\caption[Sketch of the fluorescence microscopy set-up]{\textbf{Sketch of the fluorescence microscopy set-up.} BS: Beam Splitter, C: Collimator, CamObj: Camera Objective, CryoObj: Cryo objective lens, CCD: Charge-coupled Device, F1: $785\pm3$~nm Band-pass Filter, F2: \SI{800}{\nano\meter} Long-pass Filter, GM: Galvo Mirror, HWP: Half Wave Plate, L1: Achromatic Doublet Lens, M: Mirror, P: Polariser, PD: Photo-Diode, PS: XYZ Position stage, SM: Steering Mirror, WL: White Light.}
	\label{fig:setup}	
\end{figure*}

\vspace*{\fill}

\newpage
~
\vspace*{\fill}

\begin{figure*}[h]
	\begin{center}
	\includegraphics{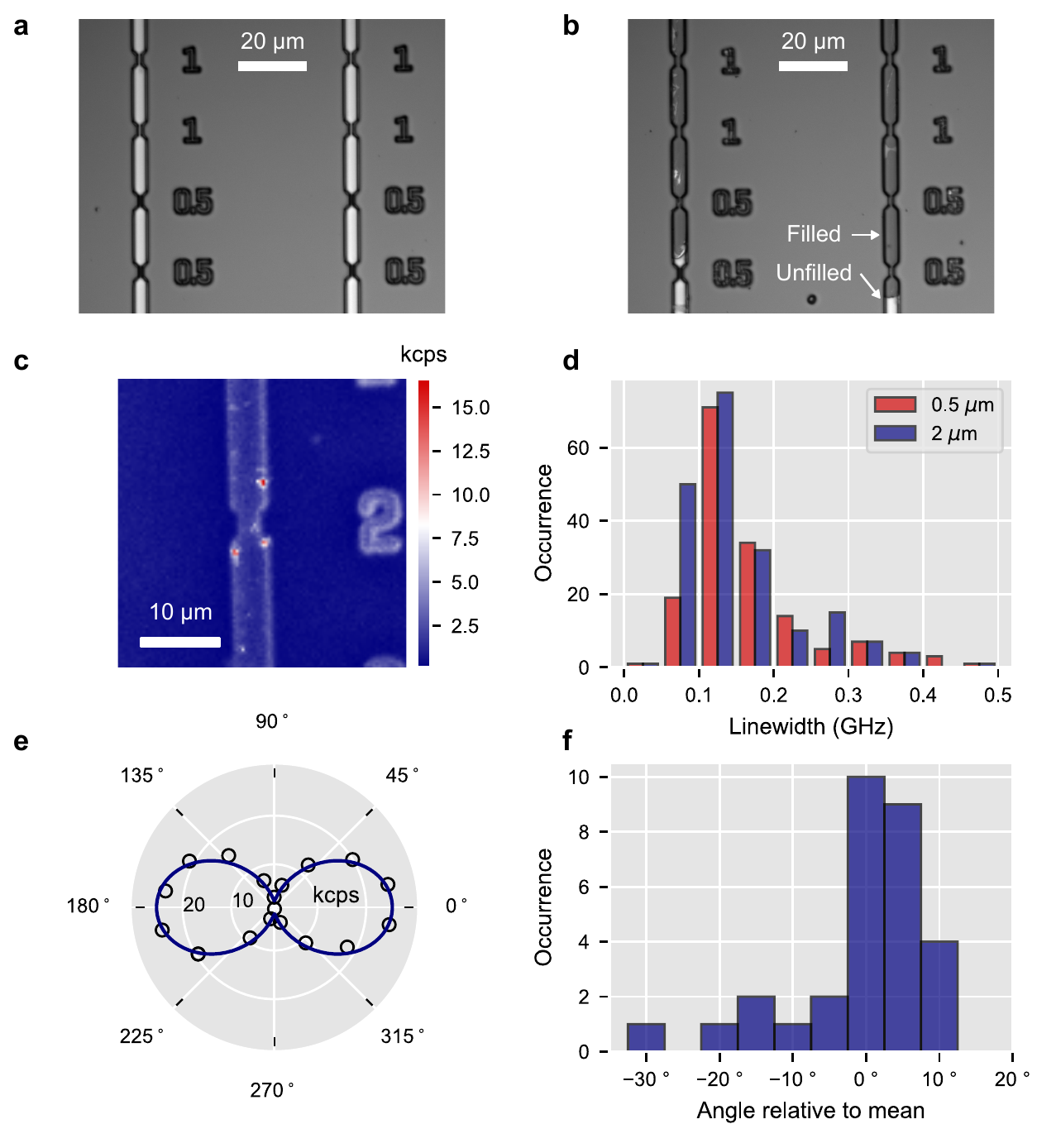}
	\end{center}
	\caption[Characterisation of filled micro-fluidic channels]{\textbf{Characterisation of filled micro-fluidic channels.} \textbf{a}, Optical microscope image of unfilled channels after removal of the sacrificial polymer structures. \textbf{b}, Image of the same channels after filling with doped anthracene. \textbf{c}, Scanning confocal fluorescence microscopy of a \SI{2}{\micro\meter} channel at \SI{4.7}{\kelvin} with the laser set at \SI{785}{\nano\meter}. \textbf{d}, Low-power linewidth distribution of DBT molecules in three \SI{2}{\micro\meter} and three \SI{0.5}{\micro\meter} channels. \textbf{e}, Fluorescence intensity at low saturation of a single molecule inside a \SI{0.5}{\micro\meter} channel as a function of excitation laser polarisation. \textbf{f}, Histogram of optical dipole moment orientations (projected into the focal plane) in a \SI{0.5}{\micro\meter} channel. The zero is the mean orientation of all the molecule angles.}
	\label{fig:cold_channel}	
\end{figure*}

\vspace*{\fill}

\end{document}